%
\magnification=\magstep1
\parskip=0.2cm
\parindent=1cm
\raggedbottom
 
\def\etal{{\it et al.\/}}
\def\pp{\parshape 2 0truecm 16truecm 1truecm 15truecm}
%
\def\apjref#1;#2;#3;#4 {\par\pp#1,  #2,  #3, #4 \par}
%
%
\def\oldapjref#1;#2;#3;#4 {\par\pp#1, {\it #2}, {\bf #3}, #4. \par}
%
%
\def\apj{ApJ}
\def\apjl{ApJLett}
\def\aap{A\&A}

\def\mnras{MNRAS}
\def\aj{AJ}
\def\araa{ARA\&A}

\def\apjs{ApJS}

%
%
\def\subsection#1{\goodbreak\noindent\underbar{#1}}
\def\subsubsection#1{{\noindent \it #1}}
\def\ltsima{$\; \buildrel < \over \sim \;$}
\def\simlt{\lower.5ex\hbox{\ltsima}}
\def\gtsima{$\; \buildrel > \over \sim \;$}
\def\simgt{\lower.5ex\hbox{\gtsima}}
%

%
\def\ed{Olszewski}

\bigskip

\centerline{\bf GALACTIC BULGES}
\medskip
\noindent{\underbar{  Rosemary  F.G.~Wyse$^1$,  Gerard 
Gilmore$^{2,3}$,  Marijn Franx$^4$}}

\medskip
\noindent $^1$ Department of Physics and Astronomy, The Johns Hopkins University,
Baltimore, MD 21218, USA

\noindent $^2$ Institute of Astronomy, Madingley Road, Cambridge CB3 0HA, UK

\noindent $^3$ Institut d'Astrophysique de Paris, 98bis boulevard Arago, 75014
Paris, France

\noindent $^4$ Kapteyn Astronomical Institute, University of Groningen, PO Box
800, 9700AV Groningen, Netherlands.

\bigskip
\noindent KEY WORDS: Galactic bulges; galaxy formation; The Galaxy;
Extragalactic astronomy; Local Group; Dynamical astronomy.

\medskip
\noindent send proofs and correspondence to: 

\item{} Dr. Gerry Gilmore, 
\item{} Institute of Astronomy
\item{} Madingley Road
\item{} Cambridge CB3 0HA
\item{} England
\item{} phone: +44-1223-337548 (reception)
\item{} phone: +44-1223-337506 (direct)
\item{} FAX: +44-1223-337523

\vfill\eject
\noindent {\bf Section 1: Motivation and Scope of Review}
\medskip

\noindent \S1.1 Introduction

In his introduction to the report of IAU Symposium \# 1, {\sl Coordination
of Galactic Research}, held near Groningen, June 1953, Blauuw noted {\it
`In the discussion the terms `halo', `nucleus' and `disk' are used to
indicate different parts of the Galaxy. These general regions are not
defined more precisely. Their introduction proved very useful, and one
might rather say that their more exact description is one of the problems
of galactic research.'} This statement provides an excellent example of the
limitations of terminology, and of the term {\sl galactic bulge} in that
this component continues to lack a clear definition (nucleus? halo?) of
either its structure or its relationship to the other stellar components of
the Galaxy.  This is compounded by the difficulty of observing bulges even
once one has decided which part of the galaxy that is.

The common usage of `bulge', for example in the term `bulge-to-disk
ratio' allocates all `non-disk' light in any galaxy which has a `disk'
into the `bulge'.  That is, the bulge contains any light that is in
excess of an inward extrapolation of a constant scale-length
exponential disk.  Sandage (Carnegie Atlas of Galaxies, [Sandage \&
Bedke 1994]; panel S11 and p45) emphasises that `One of the three
classification criteria along the spiral sequence is the size of the
central amorphous bulge, compared with the size of the disk. The bulge
size, seen best in nearly edge-on galaxies, decreases progressively,
while the current star formation rate and the geometrical entropy of
the arm pattern increases, from early Sa to Sd, Sm and Im types.'
This is the clearest convenient description of a `bulge', namely a
centrally-concentrated stellar distribution with an amorphous --
smooth -- appearance.  Note that this implicitly excludes gas, dust,
and continuing recent star formation by definition, ascribing all such
phenomena in the central parts of a galaxy to the central disk, not to
the bulge with which it cohabits.  Further, for a bulge to be
identified at all it must, by selection, have a central stellar
surface density which is at least comparable to that of the disk,
and/or it must have a (vertical) scale height which is at least not
very much smaller than that of the disk.  The fact that this working
definition can be applied successfully to the extensive
classifications in the Carnegie Atlas illustrates some fundamental
correctness.  It is also clear that bulges are very much a defining
component whose properties underly the Hubble sequence, and hence the
reason why we care -- understanding how `bulges' form and evolve is
integral to the questions of galaxy formation and evolution.

This review considers the current wide-spread beliefs and preconceptions
about `Galactic Bulges' -- for example, that they are old, metal-rich and
related to elliptical galaxies -- in the light of modern data. Our aim is
to provide an overview of interesting and topical questions, and to
emphasise recent and future observations that pertain to the understanding
of the formation and evolutionary status of `bulges'. We begin by
considering some common preconceptions.

\noindent \S1.2  Preconception Number 1: Bulges are Old. 

The expectation of `old age' arose, as far as we can ascertain, from the
interpretation of the observed correlation between stellar kinematics and
metallicity for local stars in the Milky Way by Eggen, Lynden-Bell and
Sandage (1962).  These authors proposed a model of Galaxy formation by
monolithic collapse of a galaxy-sized density perturbation, generalised to
models wherein the spheroidal components of galaxies -- including the
entire stellar mass of an elliptical galaxy -- formed stars {\it prior\/}
to the dissipational settling to a disk, and so contained the oldest stars
(e.g. review of Gott 1977).  The high central surface brightness of bulges
(and of ellipticals) also, assuming it corresponds to a high mass density,
implies a higher redshift of formation, for a fixed collapse factor of the
proto-galaxy, since at higher redshift the background density was higher
(Peebles 1989).

There clearly is an older component in the central regions of the Milky Way
Galaxy; the first real work on the bulge (or `nucleus' as it was called at
the time), used classical `halo' tracers, such as globular clusters,
RR~Lyraes and Planetary Nebulae. Of course, one must remember that `older'
is used here in the sense in which that term was used until very recently,
which meant much older than the local disk, which contains on-going star
formation.  That is, `old' means `there is no obvious AF star
population'. The Baade-era concept of `old' meant a turnoff in the
F-region, which is of course old only for a very metal-poor system (see
Sandage 1986, and the Carnegie Atlas for thorough reviews of Baade's
Population concept).  Further, the very idea of discriminating between ages
of 10Gyr and 15Gyr is a recent concept, in spite of the large fractional
difference.

Constraints on the redshift of formation of bulges can be obtained by
direct observations of high-redshift galaxies, for which morphological
information may be obtained with the Hubble Space Telescope (see \S 4).
In general it is difficult to disentagle the effects of age and metallicity
on stellar colours, even when the stars are resolved and colour-magnitude
diagrams may be examined.  The state-of-the art mean age determinations for
lower redshift bulges and disks are discussed in (\S 3), while the
interpretations of colour-magnitude diagrams are discussed in \S 2;
much ambiguity and uncertainty remains.

Implicit in the Eggen, Lynden-Bell and Sandage (1962) scenario was the
hypothesis that the Galactic bulge was simply the central region of the
stellar halo, traced at the solar neighbourhood by the high-velocity
subdwarfs.  These stars are old by anyone's definition.  Stellar haloes can
be studied easily only in the Local Group, and we discuss the stellar
populations in those galaxies in \S 2 below.

\noindent \S 1.3 Preconception Number 2: The Galactic Bulge is Super-Solar Metallicity 

This belief was strongly supported by study of late M-giants in
Baade's Window (cf. Frogel 1988), motivated by the Whitford (1978)
paper comparing the spectrum of the Milky Way bulge to that of the
integrated light of the central regions of external bulges and giant
Elliptical galaxies (see Whitford 1986 for a personal interpretation
of his research).  Whitford's investigation aimed to determine whether
or not the bulge of our galaxy was `normal', i.e. the same as others.
Whitford was apparently influenced, as were most people at that time,
by the interpretation of the colour-magnitude relation of Faber (1973)
to assume that bulges and ellipticals were differentiated only by
luminosity, which determined the metallicity, and that ages were
invariant, and {\it old}, with a turnoff mass of $\sim 1$M$_\odot$
(Faber 1973), at least for the dominant population.  In this case, the
most metal-rich stars in a lower luminosity bulge, like that of the
Milky Way, could be used as a template for the {\it typical\/} star in
a giant elliptical.

Whitford (1978) concluded from his data that indeed `the strengths of
the spectral features in the sampled areas of the nuclear bulge of the
Galaxy are very close to those expected from measures on similar areas
of comparable galaxies.'  However, Whitford's data were, by current
standards, of low spectral resolution, and were limited to the
following: spectra, with a resolution of 32\AA\ in the blue, and
64\AA\ in the red, for 3 regions in Baade's window and for the central
regions of five edge-on spirals of type Sa to Sb; lower spectral
resolution data for the central regions of M31; partial data - blue
wavelengths only - for one elliptical (NGC3379, E1), and full
wavelength coverage spectra for one other elliptical (NGC4976, E4)
which he emphasised did not match the Milky Way, and was anomalous.
Further, the data for Baade's window in the blue wavelength region --
where direct comparison with a `normal' elliptical galaxy was possible
-- were emphasised to be very uncertain, due to the large corrections
for reddening and foreground (disk) emission.  Thus, while the
Whitford paper was deservedly influential in motivating comparison
between stars in the Milky Way bulge and the integrated population of
external galaxies, its detailed conclusions rest on rather poor
foundations.

The results of Rich (1988), based on his low-resolution spectra, that
the mean metallicity of K/M giants in Baade's Window was twice the
solar value was very influential and widely accepted; however, it is
now apparent that line-blending and elemental abundance variations
contributed to a calibration error.  We discuss below the current
status of the metallicity--luminosity relation for bulges and for
ellipticals, and the detailed chemical abundance distribution for
stars in the bulge of the Milky Way; it is clear that while there do
exist super-metal-rich stars in the bulge of the Milky Way, they are a
minority, and their relationship to the majority population (are they
the same age?) remains unknown.

\noindent \S 1.4 Preconception Number 3: Bulges are similar to elliptical galaxies

Bulges and ellipticals have traditionally been fit by the same surface
brightness profiles, the de~Vaucouleurs $R^{1/4}$ law; one is tempted for
simplicity to assume that bulges are simply scaled-down ellipticals, and
that they formed the same way.  N-body simulations (e.g.  van Albada 1982),
together with analytic considerations of `maximum entropy'
endstates (Tremaine, Henon and Lynden-Bell 1986) suggested that this
was through violent relaxation of a dissipationless, perhaps lumpy, system.
These ideas incorporate the proposition (e.g. Toomre 1977; Barnes and
Hernquist 1992) that equal-mass mergers destroy pre-existing stellar disks,
and form bulges and ellipticals, these latter two being distinguished only
by mass.

Further, the stellar kinematics of ellipticals and bulges of the same
luminosity are similar, in that each rotates about as rapidly as predicted
by isotropic oblate models (Davies {\it et al.} 1983).  However, it is
becoming clear that each of `bulges' and `ellipticals' is a somewhat
heterogeneous classification, and may cover systems that formed in a
variety of ways, as discussed below. 

\medskip

The above preconceptions may be tested against modern data.  We proceed
with the systems for which the most detailed data may be obtained, the
galaxies in the Local Group, and then outward in distance.

\bigskip

\centerline{\bf Section 2 : Resolved Bulges --  Local Group Galaxies}
\bigskip

The Local Group provides a sample of bulges in which one can determine
the stellar distribution functions on a star-by-star basis, allowing a
more detailed analysis than is possible based on the integrated
properties of more distant bulges/haloes.  In this comparison, one
must be careful to isolate the essential features, since there is much
confusing detail, both observational and theoretical, specific to
individual galaxies.

Obvious questions which can be addressed most efficiently locally
include possible differences or similarities or smooth(?) gradients in
properties -- kinematics, chemical abundance distribution, age
distribution, scale-lengths, profiles etc -- from inner `bulges' to
outer `haloes', and from `bulges' to inner disks.  Different tracers
can be used, allowing comparisons between e.g. globular clusters and
field stars.

\noindent {\S 2.1 \bf Milky Way Galaxy}

Let us adopt for the moment the working definition of `the bulge' as
the component constituting the amorphous stellar light in the central
regions of the Milky Way.  While one might imagine that the Milky Way
bulge can be studied in significantly more detail than is possible in
other galaxies, our location in the disk restricts our view such that
this is true only several kpc from the Galactic centre.  Most of the
Galactic bulge is obscured by dust and stars associated with the
foreground disk. We illustrate the situation in Figure~1 below.

{\bf Figure 1 here: a full page, landscape image}

\noindent \S 2.1.1 Chemical Abundances

Chemical abundances of K- and M-giants in the central regions of the
Galaxy have been determined by a variety of techniques, ranging from
high-resolution spectra allowing elemental abundance analyses, to
intermediate-band photometry.  Application to Baade's window --
approximately 500~pc projected distance from the Galactic center --
determined that the metallicity distribution function (calibrated onto
a [Fe/H] scale) of K/M giants is broad, with a maximum at $\sim
-0.2$~dex (i.e. $\sim 0.6$ of the solar iron abundance) and extending
down to at least $-1$~dex and up to at least $+0.5$~dex (e.g.
McWilliam and Rich 1994; Sadler, Rich and Terndrup 1996). It remains
unclear to what extent these upper and lower limits are a true
representation of the underlying distribution function and to what
extent they are observational bias, set by calibration difficulties
and/or sensitivities of the techniques.  Further, the identification
of foreground disk stars remains difficult.

At larger Galactocentric distances, Ibata and Gilmore (1995a,b)
utilised fibre spectroscopy down many lines-of-sight to mimic
`long-slit spectroscopy' of the Galactic Bulge, in order to facilitate
a direct comparison between the Milky Way bulge and those of external
spiral galaxies.  They obtained spectra of $\sim 2000$ stars; star
count models, stellar luminosity classifications and kinematics were
used to isolate $\sim 1500$ K/M-giants from 700~pc to 3.5~kpc
(projected distance) from the Galactic Center.  These authors
estimated metallicities from the Mg`b' index, calibrated against local
field stars; thus there is a possible zero-point offset of up to $\sim
0.3$~dex, dependent on the element ratios of the Bulge stars compared
to the local stars.  They truncated their distribution function above
the solar value, due to the great similarity in low-resolution spectra
between foreground K dwarfs and such metal-rich K giants, which leads
to an inability to identify contamination of the bulge sample by disk
stars.  The find that the outer bulge metallicity distribution
function peaks at $\sim -0.3$~dex, and continues down beyond $-1$~dex
(see Figure~2 below).

Minniti {\it et al.\/} (1995) present the metallicity distribution
function for $\sim 250$~K/M-giants in two fields at projected
Galactocentric distances of $R \sim 1.5$~kpc.  Their results are
calibrated only for stars more metal-poor than $\sim -0.5$~dex, and
one of their fields was selected with a bias against high
metallicities.  Their data for their unbiased field again shows a
broad distribution function, approximately flat from $-1$~dex to
$+0.3$~dex.  Minniti {\it et al.\/} (1995) also summarize (and list
the references to) results from extant photometric chemical abundance
determinations (e.g. Morrison and Harding 1993); in general these
agree neither with each other, nor with spectroscopic determinations.
Further work is clearly needed.

The few large scale kinematic surveys of the bulge (Ibata and Gilmore
1995a,b; Minniti {\it et al.\/} 1995) find no convincing evidence for
an abundance-kinematics correlation within the bulge itself, after
corrections for halo stars and disk stars (see also Minniti 1996).

The most striking aspects of the metallicity distribution function of
K/M giants in the bulge are its width, and the fact that there is
little if any radial gradient in its peak (modal) value, when one
considers only spectroscopic determinations.  Further data are
required to determine whether or not the wings of the distribution are
also invariant.  Certainly the very late spectral-type M-giants have a
significantly smaller scale-height than do the K-giants (Blanco and
Terndrup 1989), a fact that could be a manifestation of either a
metallicity gradient in the high-metallicity tail of the distribution
function, or of an age gradient, with a small scale-height,
metal-rich, younger population being concentrated to the Galactic
Plane.  It is clear that star formation occurs in the very center of
the Galaxy (e.g.  Gredel 1996) so that the meaning of a distinction
between inner disk and bulge stellar populations remains problematic,
and perhaps semantic, in the inner few hundred parsecs of the Galaxy.
External disk galaxies do show colour gradients in their bulge
components, but the amplitude is luminosity-dependent and expected to
be small for bulges like that of the Milky Way (Balcells and Peletier
1994).

The little evidence there is concerning the stellar metallicity
distribution of older stars in the inner disk is also somewhat confusing.
An abundance gradient with the mean rising $\sim 0.1$~dex/kpc towards the
inner Galaxy, but for data only relevant to Galactocentric distances of
4--11~kpc, is plausibly established for F/G stars of ages up to $10^{10}$yr
(Edvardsson \etal\ 1993; their Table~14 -- their few older stars show no
evidence for a gradient).  A similar amplitude of metallicity gradient is
seen in open clusters older than 1~Gyr, but for data exterior to the solar
circle (e.g.  Friel 1995). Earlier data for K giants however suggests no
radial abundance gradient, with a mean [Fe/H]$ \sim -0.3$ from exterior to
the Sun to within 1~kpc of the center (Lewis and Freeman 1989), even though
such stars should be no older than the F/G sample. Clearly, however, the
abundance range which contains most of the bulge stars overlaps that of the
disk, with probable disk gradients being smaller than the range of the
bulge metallicity distribution function.  This is of especial interest
given the correlations, discussed below, between the colours of bulges and
inner disks in external galaxies (de Jong 1996; Peletier and Balcells
1996).

As discussed further below, the mean metallicity of field bulge stars is
significantly above that of the globular cluster system of the Milky Way,
even if only the inner, more metal-rich `disk' globular clusters, with mean
metallicity of $\sim -0.7$~dex (e.g. Armandroff 1989) are considered.

{\bf Fig 2 here}

A characterisation of the width of the metallicity distribution comes
from the fact that the distributions for both Baade's window (Rich
1990) and for the outer bulge (Ibata and Gilmore 1995b) are consistent
with the predictions of the `Simple Closed Box' model of chemical
evolution.  This is in contrast to the disk of the Milky Way, at least
in the solar neighbourhood, which has a significantly narrower
metallicity distribution, and indeed a shortage of low-metallicity
stars compared to this model (the `G-Dwarf problem').  This of course
does {\it not\/} mean that any or all of the assumptions inherent in
the simple closed box model were realised during bulge formation and
evolution, but is rather a way of quantifying the greater { width\/}
of the observed metallicity distribution in the bulge compared to the
disk at the solar neighbourhood, two locations which have the same
{\it mean\/} metallicity.

Elemental abundances provide significantly more information than do
`metallicity' since different elements are synthesised by stars of
different masses and hence on different timescales (e.g. Tinsley
1980). Different scenarios for the formation of the bulge could in
principle be distinguished by their signatures in the pattern of
element ratios (Wyse and Gilmore 1992).  The available data are
somewhat difficult to interpret, in part due to small number
statistics (e.g. McWilliam and Rich 1994; Sadler, Rich \& Terndrup
1996) but this can be rectified with the coming 8--10m class
telescopes.

\noindent \S 2.1.2 Age Estimates

RR~Lyrae stars, traditional tracers of an old, metal-poor population,
are found in significant numbers along `bulge' lines-of-sight, at
characteristic distances which place them close to the Galactic Center
(Oort and Plaut 1975).  This has been taken as supporting evidence for
an old bulge; indeed Lee (1992) argued that for a stellar population
of a high mean metallicity, such as that of the bulge discussed above,
to produce significant numbers of RR~Lyrae stars (from the metal-poor
tail of the chemical abundance distribution), it must be older than a
metal-poor population with the same RR~Lyrae production rate.  Lee
hence concluded that the bulge contained the oldest stars in the
Galaxy, older than the stars in the field halo.  But are the observed
RR~Lyrae stars indeed  part of the metal-rich bulge, or of the metal-poor
stellar halo, whose density of course also peaks in the inner Galaxy?

The samples of RR~Lyraes available for this experiment have been
small.  However, a side-benefit of the recent interest in microlensing
surveys of the Galactic bulge (e.g. OGLE; MACHO; DUO) has been
well-defined catalogs of variable stars, including RR~Lyraes.  In an
analysis of the projected spatial distribution of DUO RRLyraes,
segregated statistically by metallicity based on periods, and fitting
to density laws of halo, disk, and bulge, Alard (1996) has found that
the great majority of RR~Lyrae stars in his catalog are {\it not\/}
associated with the bulge, but rather with the thick disk and halo.
Nonetheless, a detectable fraction of the most metal-rich RR~Lyrae
variables of the 1400 discovered by DUO do indeed belong to a
concentrated bulge population. These stars comprise only about
7~percent of the whole RR~Lyrae sample. Thus it is likely that the
microlensing surveys have in fact made the first discovery of true
bulge RR~Lyraes. The intermediate abundance RR Lyraes are primarily
thick disk, while the most metal poor are primarily halo, from this
analysis.

Analysis of the variable stars detected by the IRAS satellite (mostly Mira
variables) implied a significant intermediate-age population (e.g. Harmon
and Gilmore 1988), perhaps that traced by the carbon stars (Azzopardi
\etal\ 1988; Westerlund \etal\ 1991) and the strong red clump
population (e.g. Pacynski et al 1994a,b).

Renzini (1995) has emphasised that the relative strength
of the red clump and red giant branches is dependent on helium content
as well as on age, and should the bulge stars be of high helium
content -- as expected if they were super-metal-rich -- then the
observed red clump could still be consistent with an old age. However,
the fact that the mean metallicity of the bulge is now established,
from unbiased tracers, to be below the solar value, with a
correspondingly much reduced helium abundance, makes this unlikely,
and supports intermediate age as the explanation.

Understanding the effects of dust along the line-of-sight to the central
regions is crucial.  The analysis of IR data reduces some of
the reddening problems of optical data, but again the interpretation in
terms of stellar properties is far from unambiguous.  Houdashelt (1996)
concluded that a typical age of perhaps 8~Gyr, and mean metallicity
of [Fe/H]$\sim -0.3$ (adopted from the spectroscopic results discussed
above) was consistent with his IR photometry and spectroscopy for stars in
Baade's window.

Optical/near-IR colour-magnitude diagrams which extend well below the
main sequence turnoff region may be used to make quantitative
statements about mean age and age ranges of stellar populations,
modulo uncertainties in this case due to large and highly variable
extinction, to extreme crowding in the inner fields, and to the
contribution of foreground stars.  In spite of these complications
Ortolani \etal\ (1995) concluded, on the basis of a comparison of HST
colour-magnitude data for the horizontal branch luminosity functions
of an inner globular cluster and ground-based data towards Baade's
window, that the stellar population of the bulge is as old as is the
globular cluster system, and further shows negligible age range.  This
contrasts with earlier conclusions based on pre-refurbishment HST
colour-magnitude data for Baade's Window (Holtzman \etal\ 1993),
suggesting a dominant intermediate-age population.  Future improved
deep HST colour-magnitude data are eagerly awaited.

An example of the information which can be obtained is given in Figure~3, which
is a V--I, V colour-magnitude diagram from WFPC2 data (planetary camera)
obtained as part of the Medium Deep Survey (S.~Feltzing, private communication).

{\bf fig 3 here}

\medskip

\noindent \S 2.1.3  Bulge Structure

The only single-parameter global fit to the surface brightness of the
combined halo plus bulge of the Galaxy, implicitly assuming they are a
single entity, is that by de~Vaucouleurs and Pence (1978). From their
rather limited data on the visual surface brightness profile of the
bulge/halo interior to the solar Galacto-centric distance, assuming an
$R^{1/4}$-law profile, they derived a projected effective radius of
2.75~kpc, which may be de-projected to a physical half-light radius of
3.75~kpc.  As shown by Morrison (1993), the de~Vaucouleurs and Pence
density profile, extrapolated to the solar neighbourhood, is brighter than
the observed local surface brightness of the metal-poor halo, obtained from
star counts, by 2.5 magnitudes.  Since the density profile of the outer
halo is well described by a power law in density, with index $\rho(r)
\propto r^{-3.2}$, and oblate spheroidal axis ratio of about 0.6 (Larsen
and Humphreys 1994; Wyse and Gilmore 1989), this result actually provides
the first, though unappreciated, evidence that the central regions of the
galaxy are predominantly bulge light, and that the
bulge light falls off faster than does the outer halo light. That is, the
bulge and halo are not a single structural entity.
More generally, since the spatial density distribution of the stellar
metal-poor halo is well described by a power law, while the inner
bulge (see below) is well described by another power-law of much
smaller scale length, the apparent fit of the single $R^{1/4}$-law profile
must be spurious and misleading.

The limiting factors in all studies of the large-scale structure of
the stellar Galactic bulge are the reddening, which is extreme and
patchy, and severe crowding.  The systematic difference between the
best pre-HST photometry in crowded regions and the reality, as seen by
HST, is now well appreciated after many studies of globular
clusters. Near IR studies within a few degrees of the Galactic Plane
show optical extinction which has a random variation, on angular
scales down to a few arcseconds, of up to A$_V\sim 35~\rm mag$ (e.g.
Catchpole, Whitelock and Glass 1990). At southern Galactic latitudes
however, more than a few degrees from the plane, extinction is both
low (typically E$_{B-V} \sim 0.2$) and surprisingly uniform, as is
evident in the optical bulge image in Figure~1, and as exploited by
Baade.  Nonetheless, detailed star-count modelling of the inner galaxy
(Ibata and Gilmore 1995a,b; M.~Unavane private communication)
demonstrates that extinction variations are still larger, in their
photometric effects, than are the photometric signatures of different
plausible structural models.  This sensitivity to extinction, together
with the extreme crowding which bedevils ground-based photometry, is
well illustrated by the recent history of structural analyses of the
inner Milky Way disk by the OGLE microlensing group, based on low
spatial-resolution optical data.  Their initial analysis of their data
suggested that there is no inner disk in the Galaxy, only prominent
foreground spiral structure (Paczynski {\it et al.}  1994a).  After
more careful consideration of crowding, and of alternative extinction
models, this detection of  a `hole' in the disk was retracted (Kiraga {\it
et al.} 1997). The true spatial density distribution of the inner disk
remains obscure.

There are many analyses of the surface brightness structure of the
bulge, ranging from straightforward counts of late-type stars down the
minor axis (cf.  Frogel 1988 for references) through extensive 2-D
analyses (Kent, Dame and Fazio 1991), to detailed inversions of
photometric maps (e.g. Blitz and Spergel 1991; Binney, Gerhard and
Spergel 1997).  In all such cases extreme reddening near the plane
precludes reliable use of low spatial resolution data with $|b|<2$,
irrespective of the techniques used.  The zeroth order properties of
the photometric structure of the bulge are fairly consistently derived
in all such studies, and determine $\sim 350$~pc for the minor axis
exponential scale height, and significant flattening, with minor:major
axis ratio of $\sim 0.5$.  Together with a disk scale-length of around
3~kpc, this result places the Milky Way Galaxy within the scatter of
late-type disk galaxies on the correlation between disk and bulge
scale-lengths of Courteau de Jong and Broeils (1996).

Considerable efforts have been expended in the last decade to
determine the 3-dimensional structure of the Galactic bulge. These
efforts began at a serious level with analyses of the kinematics of
gas in the inner Galaxy, following the prescient work of Liszt and
Burton in the 1980s (see Liszt and Burton 1996, and Burton, Hartmann
and West 1996 for recent reviews, and introductions to the subject),
of Gerhard and Vietri (1986) together with much other work reviewed by
Combes (1991). A resurgence of interest in bar models has been
motivated in part by new dynamical analyses (e.g. Binney \etal\ 1991;
Blitz \etal\ 1993), in part by the realisation that near-IR data might
reflect the pronounced molecular gas asymmetry (Blitz and Spergel
1991), in part by gravitational microlensing results (Paczynski \etal\
1994b), and in part by the new photometric COBE/DIRBE data (Weiland
{\it et al.} 1994).

It appears that all galaxies in their central regions have non-axisymmetric
structures, often multiple structures such as bars within bars (e.g. Shaw 
\etal\ 1995; Friedli \etal\ 1996).  
The distinction between inner spiral arms, bars, lenses, local star
formation, and such like is perhaps of semantic interest except in
cases where the distortions are of large amplitude, such as to affect
the dynamical evolution. Is the Galaxy like that? The significant
question is the existence of a substantial perturbation to the inner
density distribution, and gravitational potential, associated with a
bar. Secondary questions are the shape of that bar, and its
relationship to the disk or to the bulge. The extant 3-dimensional
models of the central regions of the Milky Way derived from the COBE
surface photometry depend on systematic asymmetries of the derived
`dust-free' surface brightness with longitude of less that 0.4
magnitudes in amplitude, after statistical correction for extinction
which is locally some orders of magnitude larger in amplitude (Binney,
Gerhard and Spergel 1997).  Thus they are crucially sensitive to
reddening corrections made on a scale of 1.5~degrees (the COBE/DIRBE
resolution) although reddening varies on much smaller scales (Fig 1).

The models also provide only a smooth description of most of the known
foreground disk structure such as can be seen in Figure~1 -- the
Ophiuchus star formation region, the Sagittarius spiral arm, etc. --
and do not work at low Galactic latitudes.  This model disk must be
subtracted before bulge parameters can be derived. The best available
description of the stellar bulge derived this way suggests axis ratios
$x:y:z \sim 1.0:0.6:0.4$ (Binney, Gerhard and Spergel 1997). This is a
rather mild bar.

It is worth noting that this model, while the best currently available,
fails to explain either the high spatial frequency structure in the
photometric data, or the observed gravitational microlensing rate towards
the inner Galaxy (Bissantz, Englmaier, Binney and Gerhard
1997, in addition to having remaining difficulties with the
details of the gas kinematics in the inner Galaxy.  Further, there is
little evidence for non-axisymmetry in the potential from analyses of
stellar kinematics -- radial velocity surveys find consistency with an
isotropic oblate rotator model (e.g. Ibata and Gilmore 1995; Minniti 1996).
Although evidence for a weak bar is seen in proper-motion surveys, 
this is very dependent on the distances assigned to the stars (Zhao, Rich
and Spergel 1994).  Thus it must be emphasised that the best available
models for the inner Galaxy remain poor descriptors of the very complex
kinematics and spatial distribution of the gas (see Liszt and Burton 1996),
and of the complex kinematics of some samples of stars (e.g. Izumiura
\etal\ 1995).

Analysis of the photometric structure of the inner galaxy is a very active
field of research, promising major progress in the next few years with the
availability of the Infrared Space Observatory imaging survey data of the
inner galaxy (Perault \etal\ 1996). ISO improves on the $\sim 1^{\circ}$
spatial resolution of COBE, having typically 6arcsec resolution in surveys.
These data for the first time provide a detailed census of individual stars
and the ISM in the inner Galaxy, with sufficient resolution and sensitivity
to see single stars at the Galactic centre, thereby allowing the first ever
determination of the true 3-dimensional spatial distribution of the inner
Galaxy.

We consider the kinematics of the Galactic bulge, the halo and the
disk, and their implications for formation models, further below
(section 5).

\noindent {\bf Section 2.2 M33 (NGC 598)}

The stellar population of M~33 was reviewed by van den Bergh (1991a) where
the reader is referred for details. We discuss the significant developments
since then concerning the existence and nature of the stellar halo and
bulge. 

M33 shows photometric evidence for non-disk light, in particular in the 
central regions, but there remains uncertainty as to the nature of this light, 
and indeed whether or not there is a central bulge component, distinct from the 
stellar halo.  

Attempts to fit optical and IR data for the central regions with an
$R^{1/4}$ law generally agree with a `bulge-to-disk' ratio of only
$\sim 2\%$, or M$_{V,bulge}$ fainter than $ \sim -15$ (Bothun 1992;
Regan and Vogel 1994).  Regan and Vogel emphasise that a single
$R^{1/4}$ provided the best fit to their data. There is some evidence
from ground-based H-band imaging (Minnitti, \ed\ and Rieke 1993) and
from HST V$-$I/I CMD data (Mighell and Rich 1995) for AGB stars in the
central regions, in excess of the number predicted by a simple
extrapolation from the outer disk; these stars have been ascribed to a
young-ish centrally-concentrated bulge.  However, McLean and Liu
(1996) contend that their JHK photometry, after removal of crowded
regions, shows no resolved bulge population distinct from the smooth
continuation of the inner disk.

Is the $R^{1/4}$ component metal-poor or metal-rich? The giant branch
of the HST CMD data is consistent with a broad range of metallicity,
ranging from M15-like to 47Tuc-like, some 1.5 dex in metallicity.  The
low end of this metallicity range is consistent with that estimated
earlier from ground-based CMD data for fields in the outer `halo',
[Fe/H] $ \sim -2.2$ (Mould and Kristian 1986).  These outer fields
showed a narrow giant branch, consistent with a small dispersion in
metallicity, and thus the two datasets together are suggestive of a
gradient in the mean metallicity and metallicity dispersion.  This may
be interpreted as evidence for a centrally-concentrated more
metal-rich component, albeit following the same density profile as the
metal-poor stars.

Pritchett (1988) reported a preliminary detection of RR~Lyrae stars in
M33, again  evidence for old, probably metal-poor, stars.

The semi-stellar nucleus of M33 has a luminosity similar to that of the
brightest Galactic globular clusters, M$_V \sim -10$, and a diameter of $\sim
6$~pc. Analysis of its spectrum (Schmidt, Bica and Alloin 1990) demonstrated
that its blue colour reflects the presence of young stars (age less than 1~Gyr)
rather than extremely low metallicity; old and intermediate-age stars with
metallicity greater than 0.1 of the solar value dominate.  The relation of this
nucleus to the `bulge', if any, is unclear.

The only kinematic data for non-disk tracers in M33 are for a subset
of its $\sim 200$ `large clusters of concentrated morphology'
(Christian 1993) of which perhaps 10\% have the colours of the
classical old globular clusters of the Milky Way.  Fourteen of these
clusters have kinematics suggestive of being `halo' objects, in that
they define a system with little net rotation, and with a `hot'
velocity dispersion of order $1/\surd{2}$ times the amplitude of the
HI rotation curve (Schommer \etal\ 1991; Schommer 1993).  Estimates of
the metallicities and ages of the `populous' clusters, based on
spectrophotometry, suggest a wide range of each, with even the
`globular clusters' spanning perhaps $\sim -2$~dex to just under solar
metallicity (Christian 1993).  Improved estimates from better data are
possible and desirable.  M33 has a very large number of globular
clusters per unit field halo light, but the meaning of this is
unclear.

In summary, M33 has a low luminosity `halo', which is at least in part old and
metal poor. There is no convincing evidence for the existence of a bulge in
addition to this halo.

\noindent {\bf Section 2.3 M31  (NGC 224)}

The stellar population of M~31 was reviewed by van den Bergh (1991b) and
again we restrict discussion to significant subsequent developments.

The field non-disk population has been studied by several groups, following
Mould and Kristian (1986; see also Crotts 1986).  These authors established,
from V and I data reaching several magnitudes down the giant branch, that the
bulge/halo of M31, at 7~kpc from center, has mean metallicity like the Galactic
globular 47~Tuc, [Fe/H] $\sim -0.7$,  and a significant dispersion in
metallicity, assuming an old population, down to $\sim -2$~dex and up towards
solar.  Similar conclusions have been reached from HST data for the outer
regions of M31 ($\sim 10$~kpc) by Holland, Fahlman and Richer (1996), and by
Rich \etal\ (1996) at $\sim 30$~kpc from the centre, limiting the amplitude of
any chemical abundance gradients.

These HST data also established firmly the lack of Blue Horizontal
Branch stars in the halo of M31, confirming the suggestion by Pritchet
and van den Bergh (1987; 1988).  The horizontal branch (HB) morphology
is apparently too red for the mean metallicity, assuming that the HB
traces a population as old as the Galactic halo globular clusters, so
that the M31 field suffers a severe `second-parameter problem'.  Age
can affect HB morphology, in that younger populations are redder at a
given metallicity, other things being equal, (e.g. Lee 1993, who also
demonstrates the effects of many other parameters), so that it is of
interest to consider this possibility (while recalling that Richer
{\it et al.}  (1996) argue quite convincingly, based on relative ages
for those Galactic globular clusters with main sequence turn-off
photometry, that age is {\it not\/} the dominant `second parameter' of
HB morphology, at least in these systems).  Indeed, the presence of
bright stars, identified as intermediate-age AGB stars, has been
suggested from (pre-refurbishment) WF/PC HST VI data at least within
the inner two kiloparsecs of the bulge (Rich and Mighell 1995).
Morris {\it et al.}  (1994) argued for a ubiquitous strong luminous
AGB component, with a typical age of 5~Gyr, from their ground-based V
and I data reaching the bright giants in various fields of M31,
16--35~kpc along the major axis of the disk and one probing the halo
at 8~kpc down the minor axis (close to the field of Mould and Kristian
1986).  Rich \etal\ (1996), and also Holland, Fahlman and Richer
(1996), find no evidence for an extended giant branch in their WFPC2
HST data for fields in the outer halo, at 10--30~kpc from center, where
again the RHB/clump is dominant, with essentially no trace of a BHB.
Thus the data describing possible metallicity/age effects remain
unclear.

Large-scale surface photometry of the disk and of the bulge of M31, in
many broad-band colours, was obtained and analysed by Walterbos and
Kennicutt (1988).  They found that there was no colour gradient in the
bulge, and that the inner disk and the bulge have essentially the same
colours, being those of `old, metal-rich stellar populations'.  This
similarity of broad-band colours has subsequently been found for a
large sample of external disk galaxies, as discussed in \S 3, and
clearly must be incorporated into models of the formation and
evolution of bulges (see \S 5 below).  Walterbos and Kennicutt also
derived structural parameters for the disk and bulge that are
consistent with the correlation between scale-lengths found for the
larger sample of more distant disk galaxies by Courteau, de Jong and
Broeils (1996). In terms of total optical light, the bulge-to-disk
ratio of M~31 is about 40\%.

Pritchet and van den Bergh (1996) emphasise that a {\it single\/}
$R^{1/4}$-law provides a good fit to their derived V-band surface
photometry (from star counts), with no bulge/halo dichotomy.  The
$R^{1/4}$ component is significantly flattened, with axial ratio of
0.55, which is similar to the value for the metal-poor halo of the
Milky Way (Larsen and Humphreys 1994; Wyse and Gilmore 1989). 

In contrast to the metal-poor halo of the Milky Way, which is
apparently flattened by anisotropic velocity dispersions, the bulge of
M31 has kinematics consistent with an isotropic oblate rotator, with
mean rotational velocity of $\sim 65$~km/s and velocity dispersion of
$\sim 145$~km/s (McElroy 1983), typical of external bulges (Kormendy
and Illingworth 1982).

Thus although Baade identified the `bulge' of M31, that is, field non-disk
stars at distances up to 35kpc from the centre of M31, with `Population
II', similar to the Milky Way halo, the dominant tracers of the M31 bulge
do not share the characteristics of classical Galactic `Population II',
being neither of low mean metallicity nor having little net rotation (see
Wyse and Gilmore 1988 for further development of this point, in the context of thick disks).

There are around 200 confirmed globular clusters associated with M31
(e.g. Fusi Pecci {\it et al.}  1993).  The distribution of their
metallicities has a mean of around $-1$~dex, more metal-poor than the field
stars, with a range of perhaps one dex on either side (e.g. Huchra, Brodie
and Kent 1991; Ajhar \etal\ 1996).  The inner, metal-rich clusters form a
rapidly rotating system, while the outer metal-poor clusters have more
classical `hot' halo kinematics (e.g.  Huchra 1993; see also Ashman and
Bird 1993 for further discussion of sub-systems within the globular
clusters).  The overall globular cluster system has a projected number
density profile that may be fit by a de~Vaucouleurs profile (although the
central regions fall off less steeply) with an effective  radius of $\sim
4-5$~kpc (Battistini \etal\ 1993).  This is more extended than the
$R^{1/4}$ fit to the field stars.  Thus in terms of kinematics, metallicity
and structure, there may be evidence for a bulge/halo dichotomy in M31 if
the halo is traced by the globular clusters and the bulge by field stars.
Note that, although there are exceptions, to first order the spatial
distributions of globular cluster systems and underlying galaxy light
are similar (Harris 1991).

As seems to be case for any system studied in sufficient detail, the morphology
of the very central regions of M31 is clearly complicated, with twisted
isophotes (Stark 1977); gas kinematics which may trace a bar (e.g. Gerhard
1988); inner spiral arms (e.g. Sofue \etal\ 1994); two nuclei (Bacon \etal\ 1993)
which may indicate a tilted, inner disk (Tremaine 1995).  These phenomena
have been modelled recently by Stark and Binney (1994) by a spherical mass
distribution plus a weak prolate bar, with the bar containing one-third of the
mass within 4~kpc (the corotation radius).  The association of `the bulge' with
this bar, which one might be tempted to adopt by analogy with the Milky Way, is
unclear.

\noindent {\bf Section 2.4 LMC}

The LMC is the nearest barred galaxy, with the bar being offset from the
kinematic and isophotal centre, and embedded in an extensive disk.
A minor metal-poor old component of the LMC is seen in deep HST colour-magnitude
data (Elson, Gilmore and Santiago 1997), but its kinematics and spatial
distribution are not yet well known.
There is a significant amount of new information concerning the variable star
population of the LMC from the several microlensing experiments, in particular
for the Long Period Variables, believed to have low-mass progenitors and hence
trace older stellar populations, and RR~Lyrae variables, traditional tracers of
old, metal-poor populations.  However, most of the information has yet to be
analysed.  There has been no kinematical analysis of the LPVs since that of
Hughes, Wood and Reid (1991), who found tentative indications of classical `hot'
halo kinematics.  The old globular clusters of the LMC, against prejudice, have
kinematics consistent with being in a rotating disk (e.g. Freeman 1993).
Thus there is little evidence for a bulge or halo population in the 
LMC, except the observation that an old, metal-poor stellar population exists.

\noindent {\bf Section 2.5 General Properties of the Local Group Disk 
Galaxies}

The diversity of properties of bulges, haloes and disks evident in the four
largest disk galaxies in the Local Group is striking. The essential
properties seem to be the following. The two latest type galaxies (M33,
LMC) have no convincingly detected bulge, but both have at least some
evidence for a small population of very old, metal-poor stars. Both have
old, metal-poor globular clusters. The intermediate-type Milky Way Galaxy
contains what can be termed both a halo - metal-poor, old, extended, narrow
abundance distribution, containing globular clusters; and a bulge -
metal-rich, mostly, and perhaps exclusively, fairly old, with a very broad
metallicity distribution function, and extremely compact in spatial
scale. The earlier-type M31 has a prominent and extended bulge, which is
both quite metal-rich and fairly old, and has a broad abundance
distribution function. The only evidence for a metal-poor old halo in M31
comes from its globular clusters, and its - very few - RR~Lyrae stars and
Blue Horizontal Branch stars.  In all cases haloes are pressure supported
systems, very unlike disks, though this is perhaps as much a definition as an
observation.

Thus, while the Local Group Spiral galaxies have a definable Halo:Disk
ratio, which is apparently rather similar for all three, only the two
earlier types have a definable Bulge:Disk ratio, which is greater
for M31 than for the Milky Way.

\bigskip

\noindent {\bf Section 3 : Low Redshift, Unresolved Bulges}

\noindent \S 3.1 Bulges and Ellipticals

In the most simplified picture of galaxies, a galaxy consist of a
bulge which follows an $R^{1/4}$ profile, and an exponential disk,
while elliptical galaxies are simply the extension of bulges in the
limit of bulge-to-disk ratio tending to infinity.

The picture has been complicated by the discovery that most
intermediate luminosity ellipticals (classified from photographic
plates) have significant disks (e.g. Bender, Doebereiner and
Moellenhoff 1988; Rix and White 1990).  These disks can be very
difficult to detect, especially when seen face-on.  Kormendy and
Bender (1996) have recently proposed that ellipticals with `disky'
isophotes, which tend to be of lower luminosity than those with `boxy'
isophotes, are the natural extension of the Hubble sequence of disk
galaxies.

Futhermore, many ellipticals show nuclear disks, either from their
kinematics, or high-resolution imaging (e.g. review of de Zeeuw and
Franx 1991).  These disks are very concentrated towards the center,
and are therefore different from disks in normal spiral galaxies.
Sometimes these disks have an angular momentum vector opposite to that
of the `bulge' (e.g. IC 1459, Franx and Illingworth 1988), implying
that the gas that formed the disk did not have its genesis in the
stars of the `bulge', but was accreted from elsewhere. Notice,
however, that some spiral galaxies also show evidence for these
`nuclear disks', including the Milky Way (Genzel {\it et al.} 1996),
and the Sombrero galaxy (Emsellem \etal\ 1996).

HST observations confirm the similarity in some aspects of
low-luminosity ellipticals and bulges.  Most of these systems have
``power law'' profiles in their inner parts, with steep index
(e.g. Faber \etal\ 1996).  In contrast, most high-luminosity
ellipticals show `breaks' in their surface brightness in the inner
regions, i.e.  relatively sudden changes where the intensity profiles
flatten. It is not clear yet what formation processes have caused
these variations, although it has been suggested that the dynamical
effects of massive black holes may be responsible (Faber \etal\ 1997).

These results suggest caution in the analysis of other data, as bulges
are not necessarily the only important component near the center, and
as the formation histories of many galaxies may have been quite
different from each other.  Indeed, it is clear that the central
regions of most, if not all, galaxies contain something `unusual' --
even without the benefit of detailed HST images (e.g.  note NGC~4314
in the Hubble Atlas, which is a barred galaxy with spiral arms in the
center of the bar).

Beyond the very central regions, a systematic variation of surface
brightness profile with bulge luminosity has been established, in that
bulges in late type spiral galaxies are better fit by exponential
profiles than by the de~Vaucouleurs profile which is appropriate for
early type spirals (e.g. Andredakis, Peletier and Balcells 1995; de
Jong 1995; Courteau, de Jong and Broeils 1996).  HST imaging of late
type spirals is needed to determine better the structure of their bulges.  
Preliminary results indicate that a significant fraction of
bulges in late type spirals have power law profiles in their inner
parts (e.g. Phillips {\it et al.} 1996)

Much recent research into the properties of elliptical galaxies has
demonstrated the existence of a `fundamental plane' which
characterises their dynamical state (e.g. review of Kormendy and
Djorgovski 1989; Bender, Burstein and Faber 1993). It has also
recently been demonstrated that the bulges of disk galaxies in the
range S0--Sc (T0--T5) occupy the same general locus in this plane
(Jablonka, Martin and Arimoto 1996).  Further, these bulges have a
similar Mg2 linestrength--velocity dispersion relationship to that of
ellipticals, but with bulges offset slightly to lower linestrengths.
This offset may be due to bulges having lower metallicity, or bulges
having lower age. Contamination by disk light can produce a similar
effect.  Jablonka \etal\ argue in favour of a close connection between
ellipticals and bulges.  Balcells and Peletier (1994) find that bulges
follow a colour--magnitude relationship similar to that of
ellipticals, but that bulges have a larger scatter, and further bulges
and ellipticals of the same luminosity do not have the same colours,
but bulges are bluer; the offset is similar to that seen by Jablonka
{\it et al.} in the strength of the magnesium index, but Balcells and
Peletier interpret it as indicating a real, though complex, difference
between bulges and ellipticals.  In addition to the data noted above
on the central parts of bulges, Balcells and Peletier (1994) find that
the amplitude of radial colour gradients also varies systematically
with bulge luminosity. They interpret their results as consistent with
bright bulges ($M_R < -20$) being similar to ellipticals (despite the
colour zero-point offset) while faint bulges are perhaps associated
with disks.

{\bf Figure 4 here}

The potential well of the outer regions of disk galaxies is clearly
dominated by dark matter, while the properties of dark matter haloes
around elliptical galaxies are less well known (e.g. de Zeeuw 1995).
How do properties of bulges scale with dark haloes? Figure~4 shows the
ratio of bulge dispersion divided by the circular velocity of the halo
(derived from rotation of tracers in the disk) against bulge-to-disk
ratios. The square on the right represents  elliptical galaxies,
derived from models by Franx (1993) which assume a flat rotation
curve. The triangle on the left corresponds to the inner regions of pure
disks, as derived for a sample of Sa--Sc galaxies by Bottema (1993; it
should be noted that the inner regions of disks are not `cold', but
warm). Bulges may be seen to lie on a rather smooth sequence between these two
extreme points. This suggests that the bulges in galaxies with low
bulge-to-disk ratios may have been formed as the same time as the
disk, whereas bulges in galaxies with large bulge to disk ratios are
so much hotter than the disk that it is more likely that they formed
separately. More and better data would be valuable to improve the
diagram.

\noindent {\S3.2 Bulges and Disks}

Astronomical gospel declares that bulges are red, and disks are
blue. This is generally presumed to be derived from studies of nearby
bulges.  Unfortunately, there are very few data on which these rather
strong statements are based. The observations were difficult to do
before the advent of CCD cameras, and have been lacking afterwards
until very recently -- perhaps because the problem was considered to
be solved. Full two dimensional imaging is needed for accurate
bulge/disk decomposition, and for exclusion of dusty areas, and large
surveys with multicolour information are still rare.  Notable
exceptions are the recent studies of the colours of `normal' spiral
galaxies by de Jong (1995; de Jong 1996) and by Balcells and Peletier
(1994; Peletier and Balcells 1996).

A relationship between bulges and disks is seen clearly in
their colours. Figure~5 shows the colour of the disks, measured at two disk
scale-lengths, against the colour of the bulges, measured at half the bulge
effective radius, in the sample of Peletier and Balcells, which consists of
luminous ($M_R \simlt -21$) nearby disk galaxies, spanning the range
S0--Sbc. 

{\bf Figure 5 here}

It is noticable how large the colour range is for the bulges - almost
as large as the range of colours for the disks. Furthermore, although
some bulges are quite red, it is also clear that blue bulges exist,
and also that red disks exist.  The sample of de Jong (1996) includes
the later morphological types of disk galaxies (Sc and Sd) and shows a
similar relationship between the colours of bulge and inner
disks. These data show that there is little support for sweeping
statements such as `bulges are red, and disks are blue'.  Colour data
for the `hidden' disks in elliptical galaxies would be very
interesting.

Further, the similarity in colour between inner disk and bulge has been
interpreted as implying similar ages and metallicities for these two
components, and an implicit evolutionary connection (de Jong 1996;
Peletier and Balcells 1996). Given the difficulties of disentangling the
effects of age and metallicity even with resolved bulges, any quantification
of `similar' must be treated with caution (see Peletier and Balcells who
derive an age difference of less than 30\%, assuming old populations with
identical metallicities). We notice in passing that the ages of ellipticals
have not been determined yet to high accuracy; measurements of various
absorption linestrengths have been interpreted to indicate a wide range of
ages of the central regions of ellipticals, with no correlation between age
and luminosity (Faber {\it et al.} 1995), but this is far from
rigourously established.

A close association between bulges and disks has been suggested by
Courteau, de Jong and Broeils (1996), on the strength of a correlation
between the scalelengths of the bulge and disk; they find that bulges
have about one-tenth the scale-length of disks.  This correlation
shows considerable scatter, especially for earlier galaxies of type
Sa, and relies upon the ability to measure reliably bulge scalelengths
which are a small fraction of  the seeing.  More and better data are
anticipated.

\noindent \S 3.3 Bulges in formation at $z < 0.1$ ?

A few exceptional systems locally are candidates for young bulges.  It
is clear that gravitational torques during interactions can act to
drive gas to the central regions (e.g. Mihos and Hernquist 1994),
where it may form stars, and which may, depending on the duration of
star formation and of the interaction, be heated into a bulge.
Schweizer (1990) discusses local disk galaxies with blue bulges,
presenting them as evidence for recent bulge-building in this manner.
These galaxies include NGC~5102, an S0 galaxy with a bluer bulge than
disk, and strong Balmer absorption lines in its central regions.
Classic merger remnants such as NGC 7252 are forming disks in their
central parts, implying that these galaxies may evolve into S0's, or
early type spirals (e.g., Whitmore \etal\ 1993)

A more dramatic example of gas-rich mergers is Arp 230, which shows
classical shells in the bulge component, and a young disk rich in gas,
as displayed in Figure~6 (D.~Schmininovich and J.~van Gorkom private
communication).

{\bf Figure 6 here}

\bigskip

\noindent{\bf Section 4 : High Redshift Bulges}

Direct searches for the progenitors of local bulges may be made by the
combination of statistically complete redshift surveys of the field
galaxy population, combined with photometric and especially with
morphological data. As an example, the I-band-selected CFHT redshift
survey contains galaxies out to redshifts of order unity, and may be
analysed in terms of the evolution of the luminosity function of
galaxies of different colours, presumed to correlate with
morphological type (Lilly {\it et al.} 1995).  These data are
consistent with very little evolution in the luminosity function of
the `red' galaxies, over the entire redshift range $0 < z < 1$, and
substantial evolution in the `blue' galaxies' luminosity function,
with the colour cut dividing the sample into `blue' and `red' taken as
the rest-frame colour of an unevolving Sbc galaxy.  This lack of
evolution for red galaxies may be interpreted as showing that the
stars of bulge-dominated systems -- the `red' galaxies -- were already
formed at redshifts greater than unity, corresponding to a look-back
time of greater than half of the age of the Universe, or 5--10~Gyr
(depending on cosmological parameters).

The high spatial resolution of the Hubble Space Telescope allows
morphological information. Schade {\it et al.} (1995) obtained HST
images for a subset (32 galaxies in total) of the CFHT redshift
survey, mostly `blue' galaxies with $z > 0.5$. They found, in addition
to the `normal' blue galaxies with exponential disks and spiral arms
and red bulge-dominated galaxies, a significant population of high
luminosity ($M_B < -20$) `blue nucleated galaxies', with large
bulge-to-disk ratio (B/T $\simgt 0.5$) -- could these be bulges in
formation, at lookback times of $\sim 5$~Gyr?  Small number statistics
notwithstanding, most of the `BNG' are asymmetric and show some
suggestions of interactions. Schade {\it et al.} (1996) found similar
results for a larger sample, using just CFHT images for morphological
classification, and confirmed that `red' galaxies tend to have high
bulge-to-disk ratios.

Extending these results to even higher redshifts, and hence studies of
progenitors of older present-day bulges, has been achieved by the
identification of a sample of galaxies with $z \simgt 3$ based on a simple colour
criterion that  selects systems with a Lyman-continuum break, superposed on
an otherwise flat spectrum, redshifted into the optical (e.g. Steidel {\it
et al.} 1996a,b). Ground-based spectroscopy of 23 high-redshift candidates
provided 16 galaxies at $z > 3$ (Steidel {\it et al.} 1996b). The observed
optical spectra probe the rest-frame 1400-1900\AA\ UV, and provide a
reasonable estimate of the reddening and hence dust content, and of the
star-formation rate.  The systems are inferred to be relatively dust-free,
with the extinction at $\sim 1600$\AA\ typically $\sim 1.7$mag, corresponding
to an optical reddening in the galaxies' rest-frame of 
${\rm  E(B-V) \sim 0.3}$mag. Whether the
low dust content is a selection effect, perhaps due to fortuitous observing
line-of-sight, or is a general feature of these high-redshift galaxies is
not clear.  The co-moving space density of these systems is large -- of
order half that of bright (L $>$ L$_*$) galaxies locally, suggesting that
not too many of them can be hidden. 
The star formation rates, assuming a solar neighbourhood IMF, are typically
$\sim 10$~$M_\odot$/yr. There are interstellar absorption lines due to
various chemical species; these lines may be interpreted as indicative of
gas motions in a gravitational potential of characteristic velocity
dispersion of $\sim 200$km/s, typical of normal galaxies today.  

Morphological information from optical
HST images (Giavalisco \etal\ 1996) for 19 Lyman-break candidates, of which
6 have confirmed redshifts,  show that in the rest-frame UV (1400--1900\AA)
these systems are mostly rather similar, in contrast to the wide range of
morphological types seen at lower redshifts, $z \sim 1$, discussed above. Further, the
typical $z \sim 3$ galaxy selected this way is compact, at least in the UV,
and has a half-light radius of $\sim 2$~kpc, reminiscent of present-day
bulges in the optical.  Some of these galaxies show faint surrounding
emission which could be interpreted as `disks'.  The star-formation rates
inferred from the spectra  build the equivalent of a bulge -- say
10$^{10}$M$_\odot$ -- over a few Gyr, which spans the redshift range from $1
\sim z \sim 4$.  Similar results are obtained from $z > 3$ samples derived
from the Hubble Deep Field (Steidel {\it et al.} 1996a), and for one
galaxy at a redshift of $z = 3.43$, the central regions of which  even fit 
a de Vaucouleurs
profile (Giavalisco {\it et al.} 1995).

Thus there is strong evidence that some (parts of some) bulges are
formed at $z \simgt 3$.  However, it is hard to draw definite
conclusions about all bulges on the basis of these results, because
the observations at these redshifts can be biased.  If, for example,
half of bulges form at $z \simlt 0.5$, then we would simply not
observe those at higher redshifts. At higher and higher redshifts, we
would simply be selecting older and older bulges. Our conclusions
would become strongly biased. This is very similar to the bias for
early type galaxies discussed by  van Dokkum and Franx (1996).

\bigskip

\noindent {\bf Section 5 : Formation Scenarios}

\centerline {5.1 Are Bulges Related  to their Haloes?}

Analyses of globular cluster systems in external galaxies conclude that
they are more metal-poor in the mean than is the underlying stellar light,
at all radii in all galaxies (Harris 1991).  It is worth noting that the
Milky Way is sometimes considered an anomaly here, in that the metallicity
distribution function for the (metal-poor, aka halo) globular cluster
system is not very different from that of field halo stars, with
differences restricted to the wings of the distributions (e.g. Ryan and
Norris 1991). It is important to note however that this comparison is done
in the Milky Way at equivalent halo surface brightness levels well below those
achievable in external galaxies. The higher-surface brightness part of the
Milky Way, that part appropriate to compare to similar studies in other
galaxies, is the inner bulge. As discussed above, the metallicity there is
well above that of the globular clusters. The Milky Way is typical. More
importantly, this (single) test suggests the possibility that {\it ALL}
spiral galaxies which have globular cluster systems have a corresponding field
halo, which in turn is systematically more metal-poor and extended than is
the more metal-rich, observable, bulge. 

If this is true, the Local Group galaxies are typical, and the concept of `stellar
halo' must be distinguished from that of `stellar bulge'. Additionally,
while haloes seem ubiquitous, they are always of low luminosity, and seem
generally more extended than bulges. Bulges are not ubiquitous, being found
in earlier type galaxies, and cover a very wide range of luminosities.

What is the evolutionary relationship, if any, between bulges and haloes?
The Milky Way is an ideal case to study this, since it has both bulge and halo.
We noted above that the bulge is more metal-rich, and possibly younger than the
halo. What of its dynamics?

{\bf Fig 7 here}

In the Milky Way the bulge stars do show significant net rotation,
(e.g. Ibata and Gilmore 1995b; Minniti {\it et al.} 1995) but the very
concentrated spatial distribution of these stars leads to low angular
momentum orbits. Indeed, the angular momentum (per unit mass)
distribution of the bulge is very similar to that of the stellar halo,
and very different from that of the disk (Wyse and Gilmore 1992; Ibata
and Gilmore 1995b); see Figure~7.  As discussed below, this is
suggestive of the ELS scenario, with the bulge being the central
regions of the halo, but formed with significantly more dissipation.
Further, the available estimates of the masses of the stellar halo and
bulge give a ratio of $\sim 1:10$, which is (coincidently?) about the
ratio predicted by models in which the bulge is built-up by gas loss
from star-forming regions in the halo (e.g. Carney, Latham, \& Laird 1990; Wyse
1995). The real test of this model is determination of the {\it rate}
of formation, and chemical enrichment, of the stars in each of the
halo and bulge. This is feasible, and requires good data on element
ratios (e.g. Wyse and Gilmore 1992).

\medskip
\centerline {5.2 Accretion/Merging } 

\S 5.2.1 Destruction of Disks by Mergers

The current paradigm of structure formation in the Universe is the
hierarchical clustering of dominant dissipationless dark matter;
galaxies as we see them form by the dissipation of gas into the
potential wells of the dark matter, with subsequent star formation
(e.g. Silk and Wyse 1993). The first scales to
collapse under self-gravity are characteristic of dwarf galaxies, and
large galaxies form by the merging of many smaller systems.  The
merging rate of the dissipationless dark haloes is reasonably
straightforward to calculate (e.g. Lacey and Cole 1993).
There are unfortunately many badly-understood parameters involved in
the physics of gaseous heating/cooling and star
formation, which determine how the baryonic components evolve.
In the absence of understanding, the naive separation of different
stellar components of galaxies is achieved by the following
prescription (Baugh, Cole and Frenk 1996; Kauffmann 1996) -- star
formation occurs in disks, which are destroyed during a merger with a
significantly larger companion (the meaning of `significant' being a
free parameter to be set by comparison with observations).  In such a
merger all the extant `disk' stars are re-assigned to the `bulge', the
cold gas present is assumed to be driven to the center and fuel a
burst of star formation, and a new disk is assumed to grow through
accretion of intergalactic gas.  Ellipticals are simply bare bulges,
more likely in environments that prevent the subsequent
re-accretion of a new disk -- environments such as clusters of
galaxies (e.g. Gunn and Gott 1972).  One consequence (see Kauffmann
1996) of this prescription is that late-type spirals, which have a
large disk-to-bulge ratio, should have older bulges than do
early-type spirals, since to have a larger disk the galaxy must have
been undisturbed and able to accrete gas for a longer time.  This does
{\it not\/} appear compatible with the observations discussed above.
Bulge formation is highly likely to be more complex than this simple
prescription.

\S 5.2.2 Accretion of Dense Stellar Satellites

The central regions of galaxies are obvious repositories of accreted
systems, being the bottom of the local potential well, provided the
accreted systems are sufficiently dense to survive tidal disruption while
sinking to the centre (e.g.  Tremaine, Ostriker and Spitzer 1975). Should
the accreted systems be predominately gaseous, then the situation is simply
that described by Eggen, Lynden-Bell and Sandage (1962), with the chemical
evolution modified to include late continuing infall. [It is worth noting
that late infall of gas {\it narrows\/} resulting chemical abundance
distribution functions (e.g. Edmunds 1990), and at least the Milky Way
bulge has an observed very broad distribution.] We now consider models of
bulge formation by accretion of small stellar systems.

As discussed above, the mean metallicity of the bulge is now
reasonably well-established at [Fe/H]$ \sim -0.3$ dex (McWilliam and
Rich 1994; Ibata and Gilmore 1995), with a significant spread below
$-1$ dex, and above solar.  Thus satellite galaxies that could have
contributed significantly to the bulge are restricted to those of high
metallicity. Given the fairly well-established correlation between
mean metallicity and galaxy luminosity/velocity dispersion
(e.g. Bender, Burstein and Faber 1993; Zaritsky, Kennicutt and Huchra
1994; Lee {\it et al.} 1993) only galaxies of luminosity comparable to
the bulge can have been responsible. That is, one is immediately
forced to a degenerate model, in which most of the stellar population
of the bulge was accreted in one or a few mergers, of objects like the
Magellanic Clouds, or the most luminous dwarf spheroidals (dSph).
Since the metallicity distribution of the Bulge is very broad,
significantly broader than the solar neighbourhood, a compromise model
is feasible, in which only the metal-poor tail of the bulge abundance
distribution function has been augmented by accretion of lower
luminosity satellite galaxies.  Quantification of this statement
awaits more robust measurement of the tails of the bulge metallicity
distribution function, and of appropriate element ratios.

Limits on the fraction of the bulge which has been accreted can be derived
from stellar population analyses, following the approach utilised by
Unavane, Wyse and Gilmore (1996) concerning the merger history of the
Galactic halo.  The Sagittarius dwarf spheroidal galaxy was discovered
(Ibata, Gilmore and Irwin 1994) through spectroscopy of a sample of stars
selected purely on the basis of colour and magnitude to contain
predominantly K giants in the Galactic bulge.  After rejection of
foreground dwarf stars, the radial velocities isolated the Sagittarius
dwarf galaxy member stars from the foreground bulge giants.  The technique
(serendipity) used to discover the Sagittarius dSph allows a real
comparison between its stellar population and that of the bulge. Not only
the radial velocities distinguish the dwarf galaxy, but also its stellar
population -- as seen in Figure~8 here (taken from Ibata et al. 1994),
{\bf all} giant stars redder than B$_J -$R$\simgt 2.25$ have kinematics
that place them in the low velocity-dispersion component {\it i.e.} in the
Sagittarius dwarf. This is a real quantifiable difference between the {\it
bulge} field population and this, the most metal-rich of the Galactic
satellite dwarf spheroidal galaxies.

\noindent {\bf Figure 8 here}

Further, the carbon star population of the bulge can be compared with those of
typical extant satellites.  In this case there is a clear discrepancy between
the bulge and the Magellanic Clouds and the dSph (Azzopardi and Lequeux 1992), in
that the bulge has a significantly lower frequency of carbon stars.

Thus although accretion may have played a role in the evolution of the
bulge of the Milky Way, satellite galaxies like those we see around us now
cannot have dominated.  However, accretion is the best explanation for at
least one external bulge -- that of the apparently normal Sb galaxy
NGC~7331, which is counter-rotating with respect to its disk (Prada {\it et
al.} 1996).  It should also be noted that for S0 galaxies -- those disk
galaxies that in theory have suffered the most merging -- Kuijken, Fisher
and Merrifield (1996) have completed a survey for counter-rotating
components in the disks, and find that only 1\% of S0 galaxies contain a
significant population of counter-rotating disk stars. This is a
surprisingly low fraction, and suggests some caution prior to adopting late
merger models as a common origin of early-type systems.

\medskip
\centerline {5.3 Disk--Bars--Bulges etc}

Recall that the broad-band colour distributions of disk galaxies show
smooth continuity across the transition between disk and bulge. In the
mean there is approximate equality between the colours of the inner
disk and the bulge in any one galaxy (de Jong 1996; Peletier and
Balcells 1996).  These data may be interpreted as showing similar mean
age and metallicity for inner disk and bulge (de Jong 1996; Peletier
and Balcells 1996), but the degeneracy of age and of metallicity on
the colours of stellar populations cause uncertainties.  Courteau, de
Jong and Broeils (1996) find further that the scale-lengths of disk
and bulge are correlated, and argue that this relationship implies
that the bulge formed via secular evolution of the disk. In principle
this is possible, if disks are bar-unstable, and bars are themselves
unstable, and very significant angular-momentum transport is feasible.

The secular evolution of collisionless stellar disks has been studied
in some detail recently, in particular through three-dimensional
N-body simulations (Combes {\it et al.} 1990; Raha {\it et al.} 1991;
see Combes 1994 and Pfenniger 1993 for interesting reviews).  These
simulations demonstrated that not only are thin disks often unstable
to bar formation, but bars themselves can be unstable, in particular
to deformations out of the plane of the disk, perhaps leading to
peanut-shaped bulges.  The kinematics of stars in peanut bulges lends
some observational support for the association of peanut-bulges with
bars (Kuijken and Merrifield 1995).  Thus stars initially in the inner
disk end up in the bulge, providing a natural explanation for the
continuity observed in the properties of the stellar populations in
disks and in bulges.

Merritt and Sellwood (1994; see also Merrifield 1996) provided a
detailed description of the physics of instabilities of stellar disks.
They demonstrate that the buckling instability of the stellar bar that
produces a `peanut bulge' (Combes {\it et al.} 1990; Raha {\it et al.}
1991) is a collective phenomenon, similar to a forced harmonic
oscillator.  Thus the instability involves the bar in general, not
only stars on special resonant orbits, as had been earlier proposed
(e.g. Combes {\it et al.} 1990).  Not all instabilities form
`peanuts', which is just as well for this class of model for bulge
formation, since, while box/peanut bulges are perhaps fairly common,
comprising 20\% of galaxies (Shaw 1987), the subset of these which
rotate on cylinders is small (e.g. Shaw 1993, and refs
therein). Relevant photometric studies show that the light in a peanut
bulge is additional to that in a smooth underlying disk, not
subtracted from it (e.g. Shaw, Dettmar and Bartledress, 1990; Shaw
1993), rather weakening the case for these models.

The extant simulations of bar-instabilities also find that a very
small mass concentration can destroy the bar.  Such a mass
concentration is very likely, since inflow, driven by gravitational
torques, is probable after a bar is formed. Hasan and Norman (1990)
suggested that a sufficiently large central mass concentration could
eventually destroy the bar.  Norman, Sellwood and Hasan (1996) used
3-D N-body simulations to follow the evolution of a bar-unstable disk
galaxy, and attempt to incorporate the effects of gas inflow by
allowing the growth of a very centrally-concentrated
component. Indeed, in time the fraction of material in this central
component is sufficient to destroy the bar, fattening it into a
`bulge-like' component.  Bulges may be built up by successive cycles
of disk instability--bar-formation--bar-dissolution (Hasan, Pfenniger
and Norman 1993).  The timescales and duty-cycles are not clear.  Some
simulations (e.g. Friedli 1994) find that as little as 1\% of the mass
in a central component is sufficient to dissolve a bar.  This is a
potential problem, as Miller (1996) points out, since the fact that
one observes bars in around 50\% of disk galaxies means that bars
cannot be fragile. A numerical example supporting Miller's important
point is provided by Dehnen (1996) who finds that his bar is stable
even with a cuspy density profile in the underlying disk; the
simulations are clearly not yet mature.

A further potential problem with the general applicability of this scenario
of bulge formation is the different light profiles of bars in galaxies of
different bulge-to-disk ratio -- early-type disk galaxies have bars with
flat surface density profiles (e.g. Noguchi 1996; Elmegreen \etal\ 1996),
while late-type galaxies have bars with steeper surface brightness profiles
than their disks.  The Courteau {\it et al.}  correlation, that bulge
scale-lengths are around 1/8 that of disks, was found for a sample of 
late-type galaxies.  In this scenario the colour of a bar should also be the
same colour as its surrounding disk, so that the subsequent bulge is the
same colour as the disk.  While colours of bars are complicated by dust
lanes and associated star formation, barred structures are often identified
by means of colour maps (e.g. Quillen \etal\ 1996).

Specific counter-examples to models whereby the bulge forms through
secular evolution of the inner disk, are the high luminosity but low
surface brightness disk galaxies which have apparently `normal' bulges
(e.g. surface brightnesses and scale-lengths typical of galaxies with
high surface brightness disks), such as Malin I (McGaugh, Schombert
and Bothun 1995) that clearly could not have formed by a disk-instability.

Dissipationless formation of bulges from disks suffers yet a further
problem, in that the phase space density of bulges is too high
(Ostriker 1990; Wyse 1997).  This also manifests itself in the fact
that the spatial densities of bulges are higher than those of inner
disks. Thus one must appeal to dissipational processes to form bulges,
such as gas flows.  The presence of colour gradients in some external
bulges would support a dissipative collapse with accompanying star
formation (e.g. Balcells and Peletier 1994).  Indeed, Kormendy (1993)
has argued that many `bulges' are actually inner extensions of disks,
formed through gas inflow from the disk, with later {\it in situ }
star formation.  This complicates the interpretation of the similarity
between the colours of bulges and inner disks, which was a natural
product of a stellar-instability to form bulges from disk stars.  One
should note also that should bulges indeed not be formed at high
redshift, then dissipation is also implicated in the production of the
high spatial densities of their central regions.

It is also important to note that the term `bar' is used no less
generically than is the term `bulge'. There is a fundamental, and
rarely clarified, difference between a detectable perturbation to the
luminosity distribution, and a substantial $m=2$ perturbation to the
galactic gravitational potential.  Inspection of the delightful
pictures in the Carnegie Atlas of Galaxies suggests a continuum of
structures, with all degrees of symmetry and asymmetry (ie. $ m=1, 2,
...$) and relative amplitudes.  When is a bar fundamentally more than
the region where spiral arms meet the centre?  More importantly for
the continuing debate about the centre of the Milky Way, is it true
that ALL these structures are seen in the cold disks only? Is there
such a thing as a bar-bulge?

\medskip

\noindent {\bf Section 6:  Conclusions} 

In the Local Group, all spiral galaxies, and probably all disk
galaxies, have an old, metal-poor, spatially-extended stellar
population which we define to be a stellar halo. These seem to be the
first stars formed in what would later become the galactic potential,
though the possibility of later accretion of a minor fraction remains
viable. The bulges of  Local Group spiral galaxies are more diverse in
properties, ranging from the very luminous, intermediate metallicity
and very spatially-extended bulge of M31, through the intermediate
luminosity, centrally-concentrated, bulge of the Milky Way, to no firm
detection of a bulge in M33. 

In general, well studied bulges are reasonably old, have a near-solar
mean abundance, though importantly with a very wide abundance
distribution function, and are consistent with isotropic oblate
rotator models for their kinematics, in which the basic support is
provided by random motions, and the flattening is consistent with
additional rotational effects. Given these properties, bulges are most
simply seen as the more dissipated descendents of their haloes.

However, diversity is apparent: all bulges of disk galaxies are not
old, super-metal-rich and simply small elliptical galaxies.  This is
not to say that such systems do not exist, but that bulges are
heterogeneous.  Higher luminosity bulges seem to have a closer
affinity to ellipticals, while lower luminosity bulges prefer disks.
But even this statement does not apply to all the properties of the
stellar populations of bulges.

This diversity, together with the surprisingly limited database
available concerning the photometric, structural, and kinematic
properties of bulges, preclude firm conclusions.  Much new and much
needed data are about to become available. It will be interesting to
see if the next review on `Bulges' will actually be entitled `Disks
and Ellipticals'.

\bigskip

\noindent {\it Acknowledgements}: RFGW and GG thank NATO for a
collaborative grant. RFGW acknowledges the support of the NASA
Astrophysics Theory Program, and the Seaver Foundation, and thanks the
Berkeley Astronomy department and the Center for Particle Astrophysics
for hospitality during some of the writing of this review.

\bigskip

\centerline{\bf References }
\medskip
\parindent=0pt

\apjref Ajhar, A., Grillmair, C., Lauer, T., Baum, W., Faber, S.,
Holtzman, J., Lynds, C.R. and O'Neill, E. 1996;\aj;111;1110-1127

\pp Alard, C. 1996, PhD thesis, Universite Paris VI.

\pp Andredakis, Y.C., Peletier, R.F. and Balcells, M. 1995, MNRAS, 275, 874-888

\apjref Armandroff, T. 1989;AJ;97;375-389
\apjref Ashman, K. and Bird, C. 1993;\aj;106;2281-2290
\pp Azzopardi, M. and  Lequeux, J. 1992, in `The Stellar Populations
of Galaxies', IAU Symposium~149, eds B.~Barbuy and A.~Renzini (Kluwer,
Dordrecht) p201-206 

\apjref Azzopardi, M., Lequeux, J. and Rebeirot, E. 1988;\aap;202;L27-L29
\apjref Bacon, R., Emsellem, E., Monnet, G. and Nieto, J-L. 1994;\aap;281;691-717
\apjref Balcells, M. and Peletier, R. 1994;AJ;107;135-152
\apjref Barnes, J. and Hernquist, L. 1992;\araa;30;705-742
\apjref Battistini, P., Bonoli, F., Casavecchia, M., Ciotti, L.,
Frederici, L. and Fusi-Pecci, F. 1993;\aap;272;77-97
\pp Baugh, C.M., Cole, S. and Frenk, C.S. 1996, MNRAS, 283, 1361-1378

\apjref Bender, R., Doebereiner, S. and Moellenhoff, C. 1988;A\&AS;74;385-426
\apjref Bender, R., Burstein, D. and Faber, S. 1993;\apj;411;153-169
\apjref Bertelli, G., Bressan, N., Chiosi, C., Fogatto, F., \& Nasi,
E, 1994;A\&AS;106;275-302
\apjref Binney, J.J., Gerhard, O., Stark, A.A., Bally, J. and Uchida, K.I. 
1991;\mnras;252;210-218
\pp Binney, J., Gerhard, O., and Spergel, D. 1997, \mnras, in press

\pp Bissantz, N., Englmaier, P., Binney, J. and Gerhard, O. 1997,
MNRAS in press

\apjref Blanco, V. and Tendrup, D. 1989;AJ;98;843-852
\apjref Blitz, L. and Spergel, D.N. 1991;\apj;379;631-638
\apjref Blitz, L. Binney, J., Lo, K.Y., Bally, J. and Ho, P.T.P. 1993;Nature;361;417-424
\apjref Bothun, G. 1992;\aj;103;104-109
\apjref Bottema, R. 1993;\aap;275;16-36
\pp Burton, W., Hartman, D. and West, S., 1996 in `Unsolved Problems of the 
Milky Way', IAU Symp 169, eds L. Blitz and P. Teuben p 447-468 (Kluwer: Dordrecht)

\pp Carney, B., Latham, D., \& Laird, J.,  1990 in `Bulges of
Galaxies', eds B.~Jarvis and D.~Terndrup  (ESO, Garching) p127-135

\pp Catchpole, R., Whitelock, P. and Glass I. 1990, MNRAS, 247, 479-490

\pp Christian, C.  1993, in `The Globular Cluster -- Galaxy
Connection'. eds G.~Smith and J.~Brodie (ASP, San Francisco) p448-457

\apjref Combes, F. 1991;ARAA;29;195-238

\pp Combes, F., 1994, in `The Formation and Evolution of Galaxies',
eds C. Munoz-Turon \& F. Sanchez, (Cambridge: University Press) 317-398

\apjref Combes, F., Debbasch, F., Friedli, D. and Pfenniger, D. 
1990;\aap;233;82-95
\pp Courteau, S., de Jong, R. and Broeils, A.  1996,  \apjl, 457, L73-L76

\apjref Crotts, A. 1986;AJ;92;292-301
\apjref Davies, R.L., Efstathiou, G.P., Fall, S.M., Illingworth, G. and 
Schechter, P. 1983;\apj;266;41-57
\pp Dehnen, M. 1996, in `New Light on Galaxy Evolution', eds
R.~Bender and R.L.~Davies (Kluwer, Dordrecht) p359.

\apjref van Dokkum, P.G. and Franx, M. 1996;\mnras;281;985-1000
\apjref Edmunds, M. 1990;\mnras;246;678-687
\apjref Edvardsson, B., Andersen, J., Gustafsson, B., Lambert, D.L., Nissen,
P. and Tomkin, J. 1993;\aap;275;101-152
\apjref Eggen, O., Lynden-Bell, D. and Sandage, A. 1962;\apj;136;748-766
\apjref Elmegreen, B., Elmegreen, D., Chromey, F., Hasselbacher,
D. and Bissell, B. 1996;AJ;111;2233-2237
\pp Elson, R., Gilmore, G. and Santiago, B. 1997, \mnras, in press

\apjref Emsellem, E., Bacon, R., Monnet, G. and Poulain, P. 1996;A\&A;312;777-796
\apjref Faber, S.M. 1973;\apj;179;731-754

\pp Faber, S.M., Trager, S., Gonzalez, J., \& Worthey, G. 1995, in
`Stellar Populations', IAU Symposium~164, eds P.~van der Kruit \&
G.~Gilmore (Kluwer, Dordrecht), p249-258

\pp Faber, S.M., Tremaine, S., Ajhar, E.A., Byun, Y.-I., {\it et al.} 
1997. ApJ in press

\pp Franx, M., 1993, in `Galactic Bulges', IAU Symposium~153, eds 
H.~Dejonghe and H.~Habing (Kluwer, Dordrecht) p243-262

\apjref Franx, M. and Illingworth, G. 1988;\apj;327;L55-L59
\pp Freeman, K.C.   1993, in `The Globular Cluster -- Galaxy
Connection'. eds G.~Smith and J.~Brodie (ASP, San Francisco) p27-38

\pp Friedli, D. 1994, in 
`Mass Transfer Induced Activity in Galaxies', ed I.~Shlosman (CUP,
Cambridge) p 268-273 

\apjref Friedli, D., Wozniak, H., Rieke, M., Martinet, L. and
Bratschi, P. 1996;A\&AS;118;461-479 
\apjref Friel, E. 1995;\araa;33;381-414
\pp  Frogel, J.  1988, ARAA, 26, 51-92

\pp Fusi-Pecci, F., Cacciari, C., Federici, L. and Pasquali, A., 1993,
in  `The Globular Cluster -- Galaxy
Connection'. eds G.~Smith and J.~Brodie (ASP, San Francisco) p410-419

\apjref Genzel, R., Thatte, N., Krabbe, A., Kroker, H., \&
Tacconi-Garman, L. 1996;\apj;472;153-172

\apjref Gerhard, O.E.  1988;\mnras;232;13P-20P
\apjref Gerhard, O.E. and Vietri, M. 1986;\mnras;223;377-389
\apjref Giavalisco, M., Steidel, C. and Macchetto, D. 1996;ApJ;470;189-194
\apjref Giavalisco, M.,  Macchetto, D., Madau, P. and Sparks, B. 1995;ApJL;441;L13-L16
\apjref Gott, J.R. 1977;ARAA;15;235-266
\pp Gredel, R. (ed) 1996 `The Galactic Center' ASP Conf series, vol 102.
\apjref Gunn, J. and Gott, J.R. 1972;\apj;176;1-20
\apjref Harris, W. 1991;ARAA;29;543-580
\apjref Harmon, R., and Gilmore, G. 1988;\mnras;235;1025-1047
\apjref Hasan, H. and Norman, C. 1990;\apj;361;69-77
\apjref Hasan, H., Pfenniger, D.  and Norman, C. 1993;\apj;409;91-109
\apjref Holland, S., Fahlman, G. and Richer, H.B. 1996;\aj;112;1035-1045
\apjref Holtzman, J. {\it et al.} 1993;AJ;106;1826-1838
\apjref Houdashelt, M. 1996;PASP;108;828.
\pp Huchra, J.   1993, in `The Globular Cluster -- Galaxy
Connection'. eds G.~Smith and J.~Brodie (ASP, San Francisco) p420-431

\apjref Huchra, J., Kent, S. and Brodie, J.   1991;\apj;370;495-504
\apjref Hughes, S.A., Wood, P. and Reid, I.N. 1991;AJ;101;1304-1323
\apjref Ibata, R. and Gilmore, G. 1995a;\mnras;275;591-604
\apjref Ibata, R. and Gilmore, G. 1995b;\mnras;275;605-627
\apjref Ibata, R.,  Gilmore, G. and Irwin, M. 1994;Nature;370;194-196
\pp Ibata, R., Wyse, R.F.G., Gilmore, G., Irwin, M., and Suntzeff,
N. 1997, AJ, in press
\apjref Izumiura, H.  \etal\ 1995;\apj;453;837-863
\apjref Jablonka, P., Martin, P. and Arimoto, N. 1996;\aj;112;1415-1422
\pp de Jong, R. 1995, PhD Thesis (Groningen)
\apjref de Jong, R. 1996;\aap;313;45-64
\apjref Kauffmann, G., 1996;\mnras;281;487-492
\apjref Kent, S.,  Dame, T.M. and  Fazio, G. 1991;ApJ;378;131-138
\pp Kiraga, M., Paczynski, B. and Stanek, K. 1997, preprint

\pp Kormendy, J. 1993, in `Galactic Bulges', IAU Symposium~153, eds 
H.~Dejonghe and H.~Habing (Kluwer, Dordrecht) p209-230

\apjref Kormendy, J. and Bender, R. 1996;\apjl;464;L119-L122
\apjref Kormendy, J. and Djorgovski, S. 1989;ARAA;27;235-278
\apjref Kormendy, J. and Illingworth, G. 1982;\apj;256;460-480
\apjref Kuijken, K. and Merrifield, M. 1995;\apjl;443;L13-L16
\apjref Kuijken, K., Fisher, D  and Merrifield, M. 1996;\mnras;283;543-550
\apjref Lacey, C. and Cole, S. 1993;\mnras;262;627-649
\apjref Laird, J., Carney, B., Latham, D., \& Aguilar, L., 1988;\aj;96;1908-1917
\apjref Larsen, J. and Humphreys, R. 1994;ApJL;436;L149-L152
\apjref Lee, Y.-W. 1992;AJ;104;1780-1789
\pp Lee, Y.-W. 1993, in `The Globular Cluster -- Galaxy
Connection'. eds G.~Smith and J.~Brodie (ASP, San Francisco) p142-155

\apjref Lee, M.G., Freedman, W., Mateo, M., Thompson, I., Rath, M. and Ruiz, 
M.-T. 1993;AJ;106;1420-162
\apjref Lewis, J. and Freeman, K.C. 1989;\aj;97;139-162
\apjref Lilly, S., Tresse, L., Hammer, F., Crampton, D. and Le Fevre, O., 1995;\apj;455;108-124
\pp Liszt, H. and Burton, W.B. 1996 in `Unsolved Problems of the Milky Way', IAU
Symposium 169, eds L.~Blitz and P.~Teuben (Kluwer, Dordrecht) p297-310
 
\apjref McElroy, D. 1983;\apj;270;485-506
\apjref McGaugh, S., Schombert, J. and Bothun, G. 1995;AJ;109;2019-2034
\apjref McLean, I. and Liu, T. 1996;\apj;456;499-503
\apjref McWilliam, A. and Rich, M.,1994;\apjs;91;749-791
\pp Madsen, C. and Laustsen, S. 1986, {\it ESO Messenger} No.46, p12.

\pp Merrifield, M. 1996, in `Barred Galaxies', IAU Colloquium~157, ASP 
Conference series vol 91, eds R.~Buta, D.A.~Crocker and B.G.~Elmegreen
(ASP, San Francisco) p179-187

\apjref Merritt, D. and Sellwood, J. 1994;\apj;425;551-567
\apjref Mighell, K. and Rich, R.M. 1995;AJ;110;1649-1664
\apjref Mihos, J.C. and Hernquist, L. 1994;\apj;425;L13-L16
\pp Miller, R.H. 1996, in `Barred Galaxies', IAU Colloquium~157, ASP Conference
series vol 91, eds R.~Buta, B.G.~Elmegreen and D.A.~Crocker (ASP, San
Francisco) p569-574

\apjref Minniti, D. 1996;\apj;459;175-180
\pp Minniti, D., \ed, E. and Rieke, M. 1993, ApJL, 410, L79-L82

\apjref Minniti, D., \ed, E., Liebert, J., White, S.D.M, Hill, J.M. and
Irwin, M. 1995;\mnras;277;1293-1311
\apjref Morris, P.W., Reid, I.N., Griffiths, W. and Penny, A.J. 1994;\mnras;271;852-874
\apjref Morrison, H. 1993;\aj;106;578-590

\pp Morrison, H. and Harding, P. 1993, in `Galactic Bulges', IAU Symposium~153, eds 
H.~Dejonghe and H.~Habing (Kluwer, Dordrecht) p297-298

\apjref Mould, J.R. and Kristian, J. 1986;\apj;305;591-599

\pp Noguchi, K. 1996, in `Barred Galaxies', IAU Colloquium~157, ASP Conference
series vol 91, eds R.~Buta, B.G.~Elmegreen and D.A.~Crocker (ASP, San
Francisco) p339-348

\pp Norman, C., Sellwood, J. and Hasan, H. 1996, \apj, 462, 114-124

\apjref Oort, J. and Plaut, L. 1975;\aap;41;71-86
\apjref Ortolani, S., {\it et al.}, 1995;Nature;377;701-703

\pp Ostriker, J.P., 1990, in `Evolution of the Universe of Galaxies', ASP 
Conference Series {\bf 10}, ed R.~Kron (ASP, San Francisco), p25

\apjref Paczynski, B, {\it et al.} 1994a;\aj;107;2060-2066
\apjref Paczynski, B, {\it et al.} 1994b;ApJL;435;L113-L116

\pp Peebles, P.J.E. 1989, in `The Epoch of Galaxy Formation', eds
C.S.~Frenk {\it et al.} (Kluwer, Dordrecht) p1-14

\pp Peletier, R. and Balcells, M. 1996, \aj, 111, 2238-2242

\apjref Perault, M., \etal\ 1996;A\&A;315;L165-L168
\pp Pfenniger, D. 1993, in `Galactic Bulges',   IAU Symposium~153, eds 
H.~Dejonghe and H.~Habing (Kluwer, Dordrecht) p387-390

\apjref Phillips, A., Illingworth, G., Mackenty, J. and Franx,
M. 1996;AJ;111;1566-1574 
\pp Prada, F., Gutierrez, C.M., Peletier, R.F. and McKeith, C.D. 1996, 
\apjl, 463, L9-L12

\pp Pritchet, C. 1988, in `The Extragalactic Distance Scale', ASP
Conference series vol 4, eds C.~Pritchet and S.~van den Bergh (ASP,
San Francisco) p59-68
\apjref Pritchet, C. and van den Bergh, S. 1987;\apj;316;517-529
\apjref Pritchet, C. and van den Bergh, S. 1988;\apj;331;135-144
\pp Pritchet, C. and van den Bergh, S. 1996, in `Unsolved Problems of the
Milky Way', IAU Symp 169, eds L.~Blitz and P.~Teuben (Kluwer,
Dordrecht) p39-46

\apjref Quillen, A., Ramirez, S. and Frogel, J. 1996;ApJ;470;790-796
\apjref Raha, A., Sellwood, J., James, R. and Kahn, F.D.,
1991;Nature;352;411-412 
\apjref Regan, M. and Vogel, S. 1994;\apj;434;536-545
\pp Renzini, A. 1995, in `Stellar Populations', IAU Symposium~164, eds
P.~van  der Kruit \& G.~Gilmore (Kluwer, Dordrecht), p325-336

\apjref Rich, R.M. 1988;\aj;95;828-865
\apjref Rich, R.M. 1990;\apj;362;604-619
\apjref Rich, R.M. and Mighell, K. 1995;\apj;439;145-154
\apjref Rich, R.M., Mighell, K., Freedman, W. and Neill,
J. 1996;\aj;111;768-776 
\apjref Rich, R.M., Mould, J.R. and Graham, J. 1993;\aj;106;2252-2279
\pp Richer, H. B., {\it et al.} 1996, ApJ, 463, 602-608

\apjref Rix, H.-W. and White, S.D.M., 1990;\apj;362;52-58
\apjref Ryan, S. and Norris, J. 1991;AJ;101;1865-1879
\apjref Sadler, E., Rich, R.M. and Terndrup, D., 1996;AJ;112;171-185
\apjref Sandage, A. 1986;ARAA;24;421-458
\pp  Sandage, A. \& Bedke, J., 1994, The Carnegie Atlas of Galaxies,
Carnegie Institution of Washington Publication 638, 750pp

\apjref Santiago, B., Elson, R. and Gilmore, G. 1996;\mnras;281;1363-1374
\apjref Schade, D., Lilly, S., Crampton,
D., Hammer, F., Lefevre, O.  and  Tresse, L. 1995;\apjl;451;L1-L4
\apjref Schade, D., Lilly, S., Lefevre, O., Hammer, F. and Crampton,
D. 1996;\apj;464;79-91
\apjref Schmidt, A., Bica, E. and Alloin, D. 1990;\mnras;243;620-628
\pp Schommer, R.A., 1993, in `The Globular Cluster -- Galaxy
Connection'. eds G.~Smith and J.~Brodie (ASP, San Francisco) p458-468

\apjref Schommer, R.A., Cristian, C., Caldwell, N., Bothun, G. and
Huchra, J. 1991;\aj;101;873-883
\apjref Shaw, M. 1987;\mnras;229;691-706
\apjref Shaw, M. 1993;\mnras;261;718-752
\apjref Shaw, M., Axon, D., Probst, R. and Gately, I. 1995;\mnras;274;369-387
\apjref Shaw, M., Dettmar, R., and Bartledress, A. 1990;A\&A;240;36-51
\pp Schweizer, F. 1990, in `Dynamics and Interactions of Galaxies', ed. 
R.~Wielen (Springer-Verlag, Berlin) p60-71

\apjref Silk, J. and Wyse, R.F.G. 1993;Physics Reports;231;293-367
\apjref Sofue, Y. {\it et al.} 1994;PASJ;46;1-7
\apjref Stark, A.A. 1977;\apj;213;368-373
\apjref Stark, A. and Binney, J., 1994;ApJL;426;L31-L33
\apjref Steidel, C., Giavalisco, M., Dickinson, M. and Adelberger, K.L. 
1996a;\aj;112;352-358 
\apjref Steidel, C., Giavalisco, M., Pettini, M., Dickinson, M. and
Adelberger, K.L.  1996b;ApJL;462;L17-L21
\pp Swaters, R.,  Sancisi, R. and van der Hulst, J.M. 1996, in `New
Light on  Galaxy Evolution', IAU Symp 171, eds R.~Bender and
R.L.~Davies, (Kluwer, Dordrecht)  p450.

\apjref Tinsley, B.M. 1980;Funds Cosmic Physics;5;287-388
\pp Toomre, A. 1977, in `The Evolution of Galaxies and Stellar Populations', eds 
R.B.~Larson and B.~Tinsley (Yale Univ Obs, New Haven) p401-416

\apjref Tremaine, S. 1995;AJ;110;628-634
\apjref Tremaine, S., Henon, M. and Lynden-Bell, D. 1986;\mnras;219;285-297
\apjref Tremaine S.D., Ostriker J.P. and  Spitzer L. 1975;\apj;196;407-411
\apjref Unavane, M., Wyse, R.F.G. and Gilmore G. 1996;\mnras;278;727-736
\apjref van Albada, T. 1982;\mnras;201;939-955
\apjref van den Bergh, S. 1991a;PASP;103;609-622
\apjref van den Bergh, S. 1991b;PASP;103;1053-1068
\apjref de Vaucouleurs, G. and Pence, W. 1978;AJ;83;1163-1174
\apjref Walterbos, R. and Kennicutt, R., 1988;A\&A;98;61-86
\apjref Weiland, J.L. {\it et al.}, 1994;\apj;425;L81-L84
\apjref Westerlund, B., Lequeux, J., Azzopardi, M. and Rebeirot, E.  
1991;\aap;244;367-372
\apjref Whitford, A. 1978;ApJ;226;777-789
\apjref Whitford, A. 1986;ARAA;24;1-22
\apjref Whitmore, B. C., Schweizer, F., Leitherer, C., Borne, K, Robert, C.,
1993;\aj;106;1354-1370

\pp Wyse, R.F.G. 1995, in `Stellar Populations', IAU Symposium 164, 
eds P.~van der Kruit \& G.~Gilmore  (Kluwer, Dordrecht), p133-150
\pp Wyse, R.F.G., 1997 preprint

\apjref Wyse, R.F.G. and Gilmore, G. 1988;\aj;95;1404-1414
\apjref Wyse, R.F.G. and Gilmore, G. 1989;Comments on Ap;8;135-144
\apjref Wyse, R.F.G. and Gilmore, G. 1992;\aj;104;144-153
\apjref Wyse, R.F.G. and Gilmore, G. 1995;\aj;110;2771-2787
\apjref Zaritsky, D., Kennicutt, R. and Huchra, J. 1994;ApJ;420;87-109
\pp de Zeeuw, P.T. 1995, in `Stellar Populations', IAU Symposium 164, 
eds P.~van  der Kruit \& G.~Gilmore (Kluwer, Dordrecht), p215-226

\apjref de Zeeuw, P.T. and Franx, M. 1991;\araa;29;239-274
\apjref Zhao, H., Spergel, D.N. and Rich, R.M. 1994;\aj;108;2154-2163

\vfill\eject
\noindent {\bf FIGURE CAPTIONS}

Figure 1: An optical image of the central Galaxy, adapted from that
published by Madsen and Laustsen (1986). The field covered is
$70^{\circ} \, {\rm x} \, 50^{\circ}$.  The Galactic Plane is
indicated by the horizontal line, and the Galactic centre by the cross
in the centre of the image. Also shown is an outline of the COBE/DIRBE
image of the Galactic centre (smooth solid curve, Arendt \etal\ 1994),
an approximate outline of the Sagittarius dSph galaxy (complex curve,
from Ibata \etal\ 1997), with the four Sgr dSph globular clusters
identified as asterisks, Baade's Window (heavy circle below the
centre), the field of the DUO microlensing survey, which contains some
of the other microlensing fields (solid square, overlapping the Sgr
dSph rectangle;Alard 1996), the four fields for which deep HST
colour-magnitude data are available (open squares, near Baade's
Window), and the six fields surveyed for kinematics and metallicity by
Ibata and Gilmore (1995a,b: black/white outline boxes). The location
of Kepler's supernova is indicated as a circle, north of the
Plane. Other features of relevance include the extreme extinction,
preventing optical/near-IR low resolution observations of the bulge
within a few degrees of the Plane, and the pronounced asymmetry in the
apparent bulge farther from the Plane. The dust which generates the
apparent peanut shape in the COBE/DIRBE image is apparent. The
asymmetry at negative longitudes north of the Plane, indicated by a
large dotted circle, is the Ophiuchus star formation region, some
160pc from the Sun. The Sagittarius spiral arm contributes
significantly at positive longitudes in the Plane.

\medskip

Figure 2: Chemical abundance distribution functions, normalized to
unity, derived by Wyse and Gilmore (1995) except where noted. The
distributions are, from top to bottom, the Solar neighbourhood stellar
halo (Laird \etal\ 1988); the outer Galactic bulge (Ibata and Gilmore
1995b), truncated by them at solar metallicity; the younger stars of
the solar neighbourhood; a volume-complete sample of local, long-lived
stars; a volume-complete sample of local thick-disk stars; the
column-integral through the disk abundance distribution for the sum of
the long-lived thin disk and the thick disk.

\medskip

Figure 3a,b: HST WFPC2 Colour-magnitude data for the Galactic Bulge,
for the field at $ {\rm (l,b)=(3.6,-7)}$ identified in Fig~1 above,
from the Medium Deep Survey. The left hand panel, Fig~3a, shows the
data. Overlaid, from a by-eye fit, is a 12~Gyr isochrone for
metallicity ${\rm [Fe/H]=-0.25}$, from Bertelli etal (1994), together
with a range of other ages plotted to one side, to illustrate the
precision required, and the need for independent determinations of
extinction at each point.  The right hand panel, Fig~3b, shows the
mean line through the data, excluding extreme points, together with
the ridge line from similar HST data for the globular cluster 47
Tucanae (Santiago, Elson, and Gilmore 1996), arbitrarily offset to
match the mean line.

\medskip

Figure 4: a) The central velocity dispersion of stellar tracers,
$\sigma$, against dark halo circular velocity, $v_c$. Open symbols are
bulges, closed symbols are ellipticals. Circular velocities for the
ellipticals are derived from models, as described by Franx (1993). (b)
The ratio of velocity dispersion in the bulge to dark halo circular
velocity, $\sigma/v_c$, taken from Franx (1993), plotted as a function
of bulge-to-disk ratio, for the entire range of Hubble Type. The
triangle at left is valid for the inner regions of pure disks, the
square at right for ellipticals. Note that systems with low B/T have
kinematics almost equal to those of inner disks.

\medskip

Figure 5: The correlation between bulge colour and the colour of the
disk of the same galaxy, for the data of Peletier and 
Balcells (1996). Note that bulges are more like their disk than they
are like each other, and the very wide range of colours evident.

\medskip

Figure 6: An optical image of Arp 230, with overlaid HI contours. This
galaxy shows evidence for  shells in its outer bulge, indicating a
recent substantial accretion event, and also has a young gas-rich disk
(D.~Schmininovich and   J.~van Gorkom private  communication).

\medskip

Figure 7: Cumulative distribution functions of specific angular
momentum for the four major Galactic stellar populations. The solid
curve is the distribution for the bulge, from Ibata and Gilmore
(1995b). The other curves are taken from Wyse and Gilmore (1992): the
dashed-dotted curve represents the halo, the dotted curve represents
the thick disk, and the dashed curve represents the thin disk.  It is
clear that the halo and bulge are more like each other than they are
like the disk components.

\medskip

Figure 8: Heliocentric radial velocities of the sample of stars
observed by Ibata, Gilmore and Irwin (1994), towards $\ell= -
5^{\circ},\, b= -12^{\circ},\, -15^{\circ},\, {\rm and} \,
-20^{\circ}.$ The stars with velocities less than about 120km/s are
predominately bulge K giants. Those with velocities between about
120km/s and 180km/s are members of the Sagittarius dwarf Spheroidal
galaxy, which was discovered from this figure. Note the real
difference between the colour distributions of bulge and Sgr
members. Thus, the bulge cannot be built up by merger of several
galaxies like the Sgr dwarf. \medskip

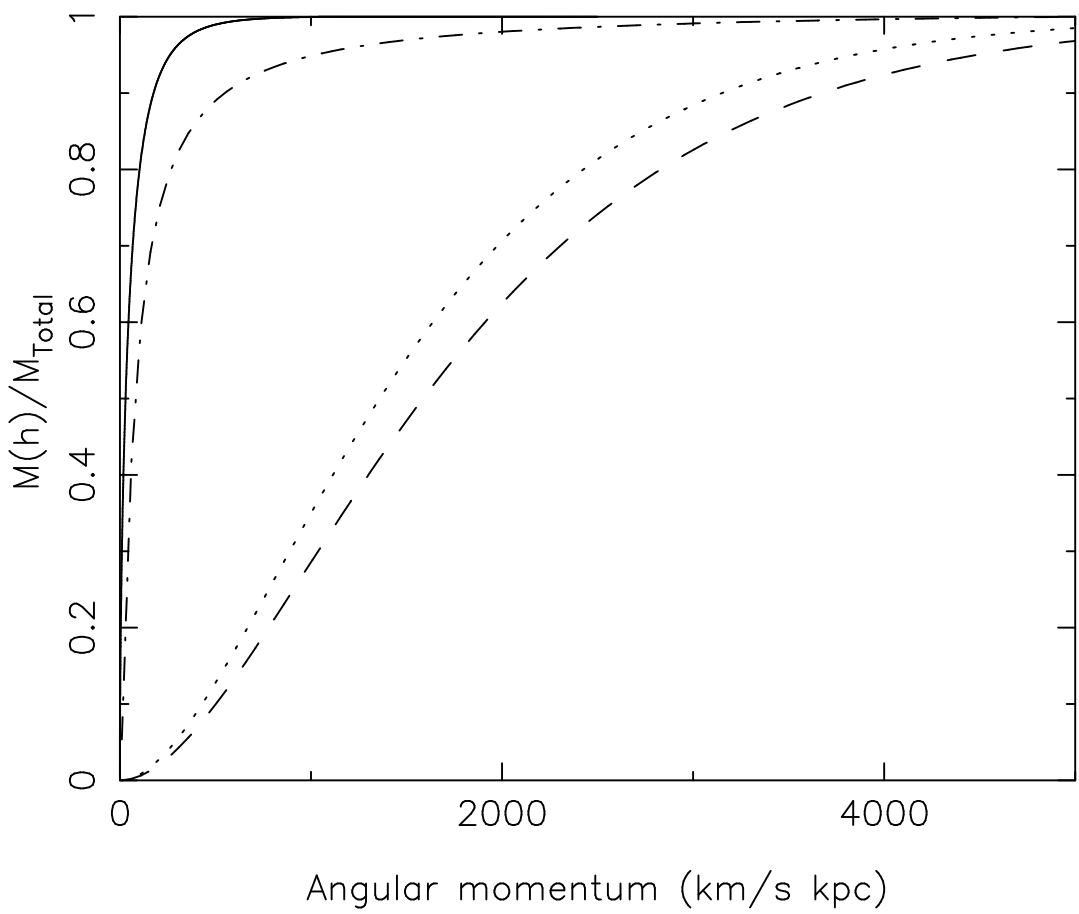
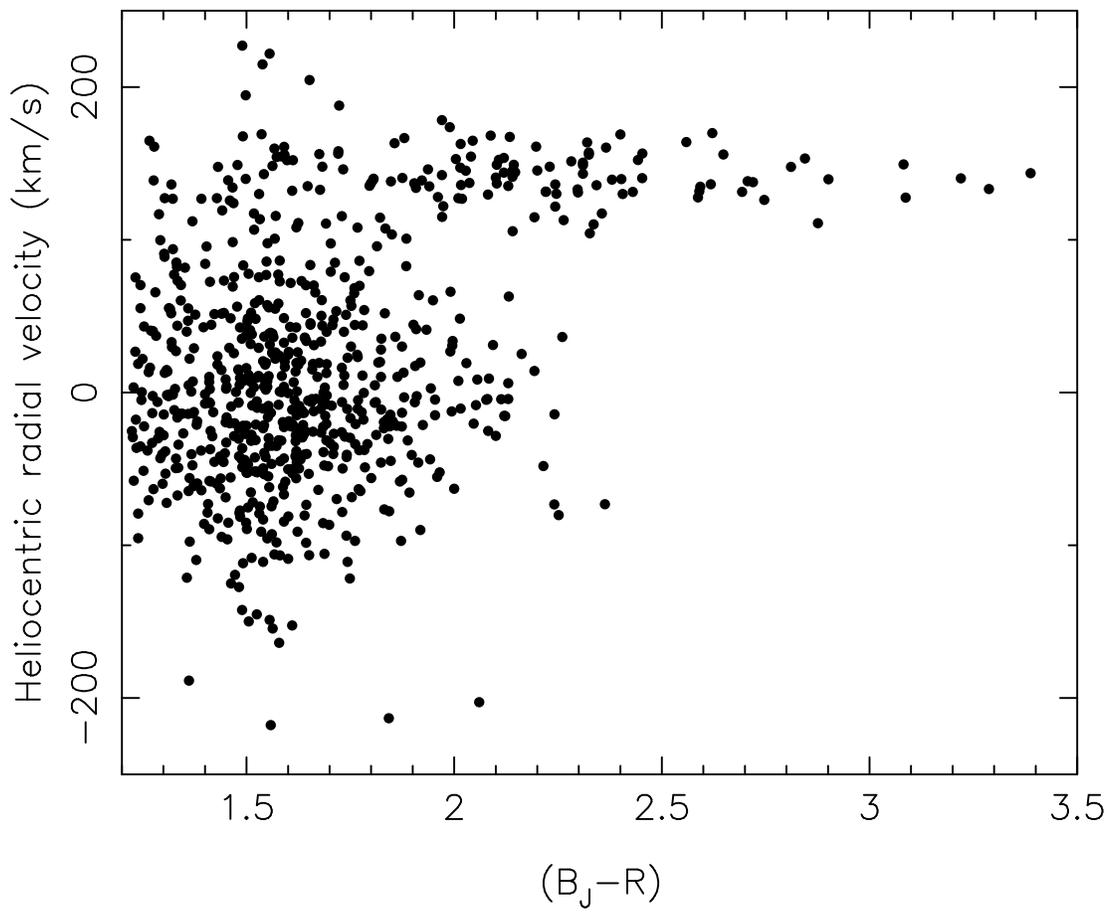
\bye